%% file: planck_nearby_galaxies_aa6.tex
\documentclass[twocolumn,traditabstract,longauth]{aa}

\input Planck.tex

\usepackage{graphicx,amsmath}
\usepackage{epsf}

\usepackage{graphicx}
\usepackage{txfonts}
\usepackage[breaklinks, colorlinks, citecolor=blue]{hyperref}
\usepackage{natbib}
\usepackage{hyperref}

\bibpunct{(}{)}{;}{a}{}{,} 

\begin{document}
\title{\textit{Planck} Early Results: The \textit{Planck} View of Nearby Galaxies}

\input{Proj_Ref_6_4a_authors_and_institutes.tex}

             \authorrunning{Planck Collaboration}
             \titlerunning{The \Planck\ View of Nearby Galaxies}
   \date{}

  \abstract
   {
The all-sky coverage of the \Planck\ Early Release Compact Source Catalogue (ERCSC)
provides an unsurpassed survey of galaxies at submillimetre (submm) wavelengths,
representing a major improvement in the numbers of galaxies detected, as well as
the range of far-IR/submm wavelengths over which they have been observed.
We here present the first results on the properties of nearby galaxies
using these data. We match the ERCSC catalogue to {\it IRAS}-detected galaxies in
the Imperial {\it IRAS\/} Faint Source Redshift Catalogue (IIFSCz), so that we can
measure the spectral energy distributions (SEDs) of these objects from
60 to $850\,\mu$m. This produces a list of 1717 galaxies with reliable associations
between \Planck\ and {\it IRAS}, from which we select a subset of 468 for SED
studies, namely those with strong detections in the three highest frequency
\Planck\ bands and no evidence of cirrus contamination.
The SEDs are fitted using parametric dust models to determine the range of dust
temperatures and emissivities. We find evidence for colder dust than has
previously been found in external galaxies, with $T<20\,$K. Such cold temperatures
are found using both the standard single temperature dust model with variable
emissivity $\beta$, or a two dust temperature model with $\beta$ fixed at 2.
We also compare our results to studies of distant submm galaxies (SMGs) which
have been claimed to contain cooler dust than their local counterparts.
We find that including our sample of 468 galaxies significantly reduces the
distinction between the two populations.  Fits to SEDs of selected objects using
more sophisticated templates derived from radiative transfer models confirm the
presence of the colder dust found through parametric fitting.
We thus conclude that cold ($T<20\,$K) dust is a significant and
largely unexplored component of many nearby galaxies.
   }


   \maketitle

\section{Introduction}

Dust is an important constituent of the interstellar medium (ISM) of galaxies.
Whilst some properties of dust in our own and very nearby galaxies can be studied
through its absorption of starlight, it was the {\it IRAS\/} satellite that first
allowed dust emission to be directly observed in large samples of external
galaxies \citep[e.g.,][]{1990ApJ...359...42D}. The all-sky {\it IRAS\/} survey at
12, 25, 60 and $100\,\mu$m in wavelength provided much new information on the
properties of dust and how this relates to other aspects of galaxies and galaxy
evolution. However, the strong temperature dependence of the spectral energy
distribution (SED) of dust emission, combined with limited wavelength coverage,
means that {\it IRAS\/} was insensitive to dust below a temperature of
$\sim30\,$K.  Observations of dust in our own Galaxy by the FIRAS instrument on
{\it COBE\/} \citep{1995ApJ...451..188R} found evidence for dust at several
different temperatures. This included a widespread component at 16--21\,K, and
another at 10--14\,K associated with molecular clouds in the inner Galaxy.
A third widespread colder component, at 4--7\,K, was later identified with the
Cosmic Infrared Background \citep[CIB;][]{1996A&A...308L...5P,1998ApJ...508..123F}.
None of these components would be detectable in
external galaxies by {\it IRAS}. {\it COBE}-DIRBE observations also detected 56
external galaxies, finding an average dust temperature of $27.6\,$K
\citep{1996astro.ph.10238O}.
  
Observations at longer far-infrared (FIR) or submillimetre (submm)
wavelengths from the ground \citep[e.g.,][]{2000MNRAS.315..115D},
from space \citep[e.g.,][]{2005ApJ...633..857D} or in
combination \citep[e.g.,][]{2009AJ....138..146W}, have provided hints that cooler
dust plays a significant role in nearby galaxies. Observations with
{\it Herschel\/} of pre-selected objects \citep[e.g.,][]{2010A&A...518L..61B} or of
relatively small fields \citep[e.g., H-ATLAS][covering up to $550\,{\rm deg}^2$]
{2010PASP..122..499E}
provide valuable data at 250 to $500\,\mu$m, which
constrain the long wavelength dust properties for specific populations.
However, the availability of the \Planck
\footnote{\Planck\ (http://www.esa.int/Planck) is a project of the European
Space Agency (ESA) with instruments provided by two scientific consortia funded
by ESA member states (in particular the lead countries France and Italy), with
contributions from NASA (USA) and telescope reflectors provided by a collaboration
between ESA and a scientific consortium led and funded by Denmark.}
Early Release Compact Source Catalogue (ERCSC) provides a long wavelength
counterpart to {\it IRAS}, allowing us an unbiased view of the FIR-to-submm
SEDs of a large, unbiased sample of nearby ($z<0.25$) galaxies. We are now, for
the first time, able to examine the role of cold dust for a wide range of objects
in the local Universe.

The discovery of the CIB \citep{1996A&A...308L...5P,1998ApJ...508..123F}
has added to the importance of our understanding of dust in galaxies. The CIB
demonstrates that roughly 50\% of all energy generated in the history of the
Universe was absorbed by dust and reprocessed into the FIR/submm
\citep{2000A&A...360....1G}. Deep surveys at $850\,\mu$m with SCUBA
\citep[e.g.,][]{1997ApJ...490L...5S,1998Natur.394..241H,
2000AJ....120.2244E,2006MNRAS.372.1621C} and at nearby wavelengths
with other instruments
(e.g., MAMBO, AzTEC and LABOCA) have revealed much higher number counts
than would be predicted by a non-evolving extrapolation of the local population.
There must thus be very rapid evolution of the FIR/submm galaxy population,
something confirmed by observations with {\it ISO},
\citep[e.g.,][]{2001A&A...372..364D} {\it Spitzer\/}
\citep[e.g.,][]{2006AJ....131..250F,2010A&A...512A..78B,2010MNRAS.tmp.1778C}
and BLAST \citep{2009Natur.458..737D}.
{\it Herschel\/} observations have now confirmed this rapid evolution through
a combination of number count
\citep{2010A&A...518L...8C,2010A&A...518L..21O} and
luminosity function \citep{2010A&A...518L..10D} studies. However, detailed
interpretation of these results is hampered by our poor knowledge of galaxy SEDs
in the 100--$1000\,\mu$m range. This is demonstrated, for example, by the apparent
separation in the temperature-luminosity plane of local {\it IRAS}-selected
galaxies and high redshift SCUBA-selected SMGs
\citep[e.g.,][]{2010MNRAS.403..274C}. The origin of this separation is unclear.
It might represent a genuine change in dust temperature with redshift, and
selection biases may be partly involved, but it could also
reflect our ignorance of the full FIR/submm SED of local galaxies.
By properly establishing a zero redshift baseline for the dust SEDs of typical
galaxies, the \Planck\ ERCSC will allow the origins of the CIB and the nature
of the galaxies that contribute to it to be much better determined.

The central goals of this paper are thus twofold: to examine the properties of
a large sample of local ($z<0.25$) galaxies to establish the range of dust
temperatures and other properties found locally; and thus to set the local
baseline against which higher redshift studies, and especially studies of the
SMGs responsible for the CIB, can be compared.
		
The rest of this paper is organised as follows. In Section~\ref{sec:Observations}
we give details of \Planck's observations of local galaxies and the ERCSC.
We also discuss the results of matching ERCSC galaxies to sources observed by
{\it IRAS}.  In Section~\ref{sec:Comparison} we present a comparison of the
ERCSC with existing data from ground-based observatories.
Section~\ref{sec:Parametric} presents the results of fitting parametric models
to the dust SEDs of ERCSC galaxies, while Section~\ref{sec:Physical}
discusses the results of physical template fitting. Finally, we draw
conclusions in Section~\ref{sec:Conclusions}.
Throughout this paper we assume a concordance
cosmology, with $H_0 = 70\,{\rm km}\,{\rm s}^{-1}{\rm Mpc}^{-1}$,
$\Omega_{\Lambda}=0.7$ and $\Omega_{\rm M} = 0.3$.
		
\section{\Planck\ Observations of Nearby Galaxies}\label{sec:Observations}

\subsection{The \Planck\ Mission}

\Planck\ \citep{tauber2010a, planck2011-1.1} is the third generation space
mission to measure the anisotropy of the cosmic microwave background (CMB).
It observes the sky in nine frequency bands covering 30--857\,GHz with high
sensitivity and angular resolution from 31\arcm\ to 5\arcm.  The Low Frequency
Instrument LFI; \citep{Mandolesi2010, Bersanelli2010, planck2011-1.4} covers the
\getsymbol{LFI:center:frequency:30GHz}, \getsymbol{LFI:center:frequency:44GHz},
and \getsymbol{LFI:center:frequency:70GHz}\,GHz bands, with amplifiers cooled to $20\,$K.  The High Frequency
Instrument (HFI; \citealt{Lamarre2010, planck2011-1.5}) covers the \getsymbol{HFI:center:frequency:100GHz}, \getsymbol{HFI:center:frequency:143GHz}, \getsymbol{HFI:center:frequency:217GHz},
\getsymbol{HFI:center:frequency:353GHz}, \getsymbol{HFI:center:frequency:545GHz}, and \getsymbol{HFI:center:frequency:857GHz}\, GHz bands, with bolometers cooled to $0.1\,$K.  Polarization
is measured in all but the highest two bands \citep{Leahy2010, Rosset2010}.  A
combination of radiative cooling and three mechanical coolers produces the
temperatures needed for the detectors and optics \citep{planck2011-1.3}.  Two
data processing centers (DPCs) check and calibrate the data and make maps of the
sky \citep{planck2011-1.7, planck2011-1.6}.  \Planck's sensitivity, angular
resolution, and frequency coverage make it a powerful instrument for Galactic
and extragalactic astrophysics as well as cosmology.  Early astrophysics results
are given in \Planck\ Collaboration, 2011h--z.

\subsection{The \Planck\ Early Release Compact Source Catalogue}

The \Planck\ ERCSC \citep{planck2011-1.10} provides positions and flux densities
of compact sources found in each of the nine \Planck\ frequency maps. The flux
densities are calculated using aperture photometry, with careful modelling of
\Planck's elliptical beams. The colour corrections for sources with spectral
index $\alpha = -0.5$ (using the convention $S_{\nu}\propto\nu^{\alpha}$) are
1.017, 1.021 and 1.030, respectively, for the
\getsymbol{LFI:center:frequency:30GHz}, \getsymbol{LFI:center:frequency:44GHz},
and \getsymbol{LFI:center:frequency:70GHz}\,GHz LFI channels.  Flux densities
taken from the ERCSC should be divided by the appropriate colour correction to
give the correct flux values for an assumed narrow band measurement at the central
frequency.
For frequencies from \getsymbol{LFI:center:frequency:30GHz} to
\getsymbol{HFI:center:frequency:143GHz} GHz, compact sources have been detected using
a version of the ``Powell Snakes'' techniques \citep{2009MNRAS.393..681C}; for
details see \cite{planck2011-1.10}.  In the four higher frequency channels,
sources were located using the SExtractor package \citep{1996A&AS..117..393B}.
Sources detected in one or more of the frequency maps were then put through a
further set of secondary selection criteria; these are discussed in detail in
\cite{planck2011-1.10}.  The primary criterion utilized was a Monte Carlo
assessment designed to ensure that $\geq$\,90\% of the sources in the
catalogue have a flux accuracy of at least 30\%.

\subsection{Matching The ERCSC to {\it IRAS\/} Data}

To understand the FIR Spectral SEDs we need a combination of data at long
wavelengths, provided by the \Planck\ ERCSC, and data near the peak of a typical
galaxy dust SED at about $100\,\mu$m. The best source for the latter information
is the {\it IRAS\/} all-sky FIR survey, and the most recent analysis of the
{\it IRAS\/} Faint Source Catalogue (FSC) is provided by
\cite{2009MNRAS.398..109W} in the Imperial {\it IRAS\/} FSC redshift survey
(IIFSCz). This was constructed using {\it IRAS\/} FSC sources, all of which are
at $|b|>20^\circ$, with {\it IRAS\/} colours used to exclude stars and cirrus
sources (with $S(100)/S(60)>8$).  The NASA Extragalactic Database (NED)
was then used to find spectroscopic redshifts for the resulting FSC source list,
and to associate the sources with SDSS
\citep[where available;][]{2000AJ....120.1579Y} and 2MASS
\citep{2006AJ....131.1163S} galaxies to
find photometry at 0.36--$2.2\,\mu$m.  This photometry was then used to estimate
photometric redshifts for sources without spectroscopic redshifts.

\begin{figure}
\centering
\includegraphics[angle=0,scale=0.4]{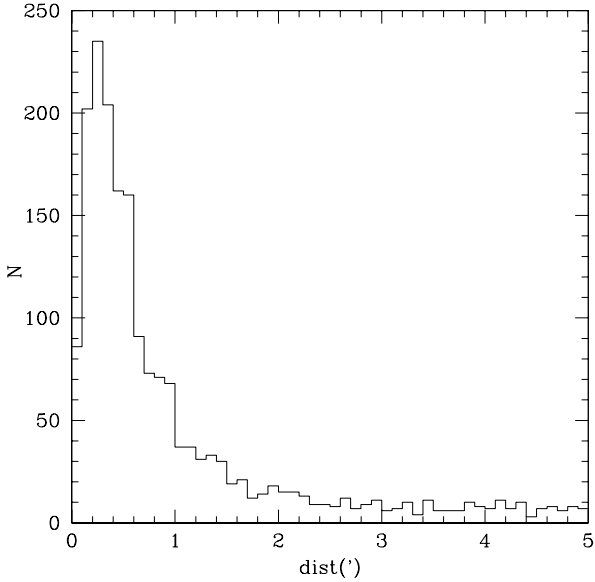}
\caption{Histogram of offsets between ERCSC and IIFSCz positions.}
\label{fig:offsets}
\end{figure}

The starting point for matching the ERCSC to the IIFSCz is the 9042 sources
detected at \getsymbol{HFI:center:frequency:857GHz}\,GHz by \Planck. This is
then restricted to the 5773 sources at $|b| > 20^\circ$ for which there will
be FSC data.
Associations of ERCSC sources with IIFSCz were looked for using a search radius
of $5^\prime$.  The histogram of positional offsets is shown in
Fig.~\ref{fig:offsets}.  The bulk of associations have offsets within $2^\prime$.
Even at $5^\prime$ there is no steep increase in the number of associations which
would be indicative of a large fraction of chance associations.  On the basis of
source surface-density, the chance of a random association with an IIFSCz source
within $3^\prime$ is 1.6\%, and within $5^\prime$ it is 4.5\%.  A total of 1966
associations were found within $5^\prime$.  There were 106 cases where an ERCSC
source picked up an association with more than one IIFSCz source.  We examined
these cases carefully to ensure that only a single association was accepted.
Generally the nearer association was preferred.  Where the offsets of the two
associations were comparable, the brighter {\it IRAS\/} source was preferred.
There are 20 cases where two bright galaxies less than $5^\prime$ apart have
generated a single ERCSC source, for which there may be a significant contribution
from both galaxies to the ERCSC flux. These confused sources would benefit from
additional observations with ground-based submm instruments to determine the
contribution of each component to the submm emission detected by \Planck. 

The remaining ERCSC-IIFSCz associations were further scrutinised as follows.
Firstly, for sources with positional offsets between the two catalogues in the
range 3--$5^\prime$ all NED associations within $5^\prime$ of the ERCSC positions
were examined to test the validity of the association with the {\it IRAS\/}
source.  This included inspection of the Sky Survey postage stamps provided in
NED.  Associations were accepted as real if the associated galaxy had a blue
($g$ or $B$) magnitude brighter than 16.  The surface-density of such galaxies
leads to the probability of a chance association being ${\sim}3$\%.  For sources
where there was both an {\it IRAS\/} and a 2MASS association, this limit was
relaxed to $B=17$ (or $K\sim13$).  Of the 88 ERCSC-IIFSCz associations with
positional offsets 3--$5^\prime$ (and with spectroscopic or photometric redshifts)
72 were associated with bright galaxies, two were associated with a second FSC
source having cirrus-like colours and are presumed to be cirrus, and the remaining
14 are classified as possible galaxy associations (these are excluded from the
reliable galaxy catalogue used here for further analysis).

The second category of ERCSC-IIFSCz  associations which we scrutinised in detail
were those for which there is no redshift in the IIFSCz.  There were 165 of these
and the NED associations suggest that 38 are bright galaxies, seven are cirrus,
two are bright planetary nebulae, and the remaining 118 are classified as
possible galaxy associations (and excluded here).  We were left with 1717
reliable ERCSC-IIFSCz galaxy associations of which 337 are flagged as extended in
the ERCSC. 1597 of these 1717 objects have spectroscopic redshifts.

\subsection{ERCSC Sources not associated with IIFSCz Sources}
\label{sec:notIIFSCz}
\begin{figure*}
\centering
\includegraphics[angle=0,scale=0.75]{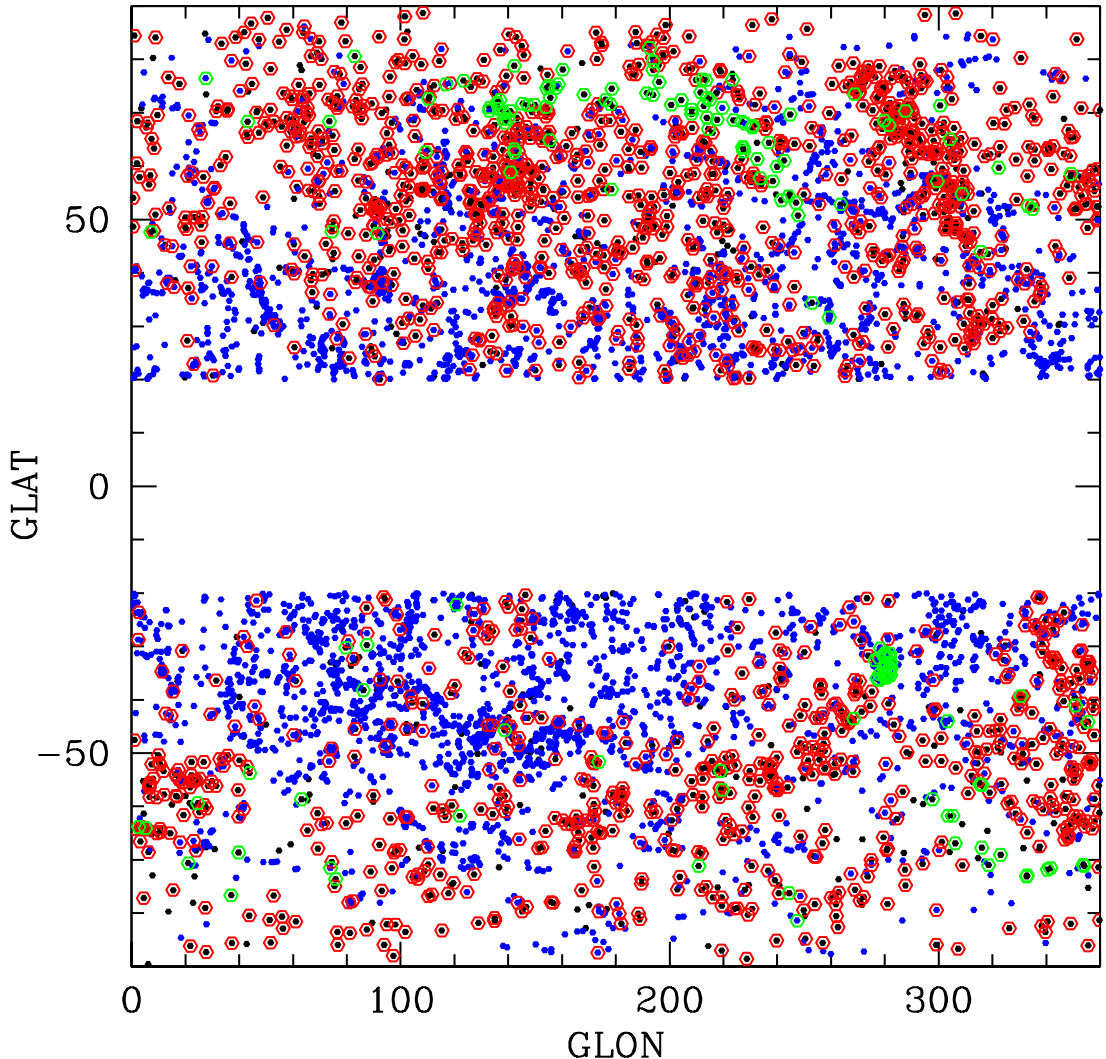}
\caption{Sky plot of ERCSC sources in galactic coordinates.  Black filled
hexagons are ERCSC point-sources and blue filled hexagons are ERCSC sources flagged
as extended.  Red hexagons are sources associated with IIFSCz {\it IRAS\/} FSC
galaxies, after scrutinising suspect categories with NED (and excluding some,
as described in the text). Green hexagons are ERCSC sources not associated with
IIFSCz, but associated with bright galaxies in NED (only for $|b|>60^\circ$ for
extended sources).
}
\label{fig:skyplot}
\end{figure*}

Fig.~\ref{fig:skyplot} shows the sky distribution of ERCSC sources at
$|b|>20^\circ$, with sources flagged as extended in the ERCSC shown as blue
filled hexagons, and point-sources shown in black.  Associations with the IIFSCz
are shown as red circles.  The extended sources not associated with IIFSCz sources
have a strikingly clustered distribution, which matches the areas of our Galaxy
with strong cirrus emission, as evidenced by {\it IRAS\/} $100\,\mu$m maps and by
the ERCSC cirrus flag (values $>0.25$).  We presume the majority of these are
cirrus sources and not extragalactic.

To test this further, we looked for NED associations with all 444 extended ERCSC
sources lacking IIFSCz associations at $|b|>60^\circ$.   Only 12 were found to
have associations with bright ($b$, $g<16$) galaxies.  Extrapolating to
$|b| = 20$--$60^\circ$, we estimate that a further $\sim50$ of these extended
non-FSC sources will be bright galaxies.  The remainder of the 3431 extended
non-FSC sources at $|b|>20^\circ$ are presumed to be Galactic cirrus.

A ridge of non-FSC point sources can be seen in Fig.~\ref{fig:skyplot} at
$b\sim70^\circ$, $l\sim120$--$230^\circ$.  These correspond to one of the
{\it IRAS\/} coverage gaps.  We examined NED associations for all 482 ERCSC
point-sources not associated with IIFSCz sources.  32 were found to be associated
with Local Group galaxies (M31, SMC and WLM, with 28 in the LMC), 123 are bright
galaxies, 27 are associated with {\it IRAS\/} FSC or PSC Galactic cirrus sources, and 10
are bright stars or planetary nebulae. Most of the bright galaxies lie in the {\it IRAS\/}
coverage gaps. The remaining 289 are classified as possible galaxy associations
(and excluded here).

To summarise, we have found a net total of 1884 definite associations with
galaxies. These constitute an ERCSC galaxy catalogue.  A further 419 sources are
not associated with bright galaxies, but there are grounds for thinking they
could be extragalactic sources.  Some have IIFSCz associations, but there are too
many possible faint optical or near-IR galaxy counterparts to be confident which
might be associated with the ERCSC source.  Some of these 419 sources are almost
certainly fainter galaxies, although many could turn out to be cirrus. Improved
submm or FIR positions are needed, either via {\it Herschel\/} or ground-based
telescopes, to identify these sources reliably.

Following this identification analysis we restrict ourselves to those galaxies
with reliable IIFSCz associations and with detections in the
\getsymbol{HFI:center:frequency:857GHz} and
\getsymbol{HFI:center:frequency:545GHz}\,GHz bands at significance of $5\sigma$ or
greater, as well as detections in the \getsymbol{HFI:center:frequency:353GHz}\,GHz
band of $3\sigma$ or greater. This amounts to a total sample size of 595 galaxies.

\subsection{Cirrus Contamination}

\begin{figure}
\centering
\includegraphics[angle=90, scale=0.37]{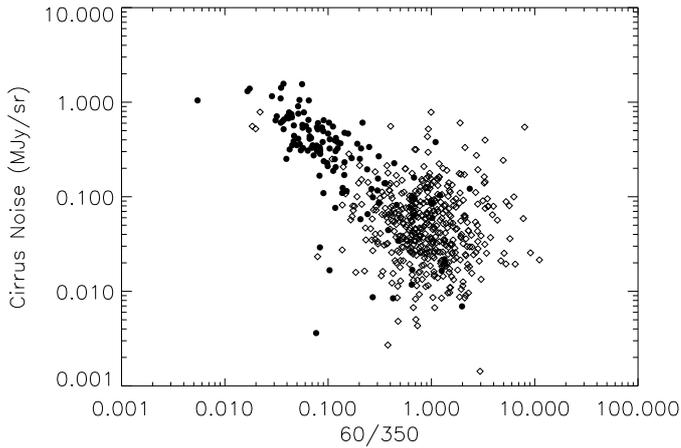}
\caption{$60\,\mu$m to \getsymbol{HFI:center:frequency:857GHz}\,GHz
(i.e., $350\,\mu$m) colour plotted against the cirrus RMS at $100\,\mu$m in
{\it IRAS}. ERCSC sources classified as point-like are shown as open diamonds,
while extended sources are shown as solid dots. Note that the extended sources
show a clear correlation between colour and cirrus RMS, indicating that these
sources are likely to be contaminated by cirrus emission.}
\label{fig:60to350}
\end{figure}

Our analysis of the non-IIFSCz-identified ERCSC sources in
Section~\ref{sec:notIIFSCz} led us to the conclusion that sources which are
extended in the ERCSC are a result of cirrus structure in our own Galaxy, or at
best are a combination of cirrus structure with flux from a galaxy.  Of the 595
reliably detected IIFSCz-identified ERCSC sources, 127 are listed as extended in
the ERCSC. We test these objects for the possibility of cirrus contamination by
examining the amplitude of the local cirrus fluctuations in the $100\,\mu$m cirrus
maps of \cite{1998ApJ...500..525S}. We adopt this approach since regions of
greatest cirrus fluctuation are those most likely to cause problems for point
source detection in the ERCSC. We measure the cirrus RMS in a $3\times3$ array
of points, separated by $0.1^\circ$ and centred on the position of the ERCSC
source. Since cirrus emission is likely to have cooler FIR-to-submm colours than
the integrated emission of an external galaxy, we then look for any correlation
between cirrus RMS and the
$60\,\mu$m-to-\getsymbol{HFI:center:frequency:857GHz}\,GHz colour. We plot this
relation in Fig.~\ref{fig:60to350}.

As can be seem from Fig.~\ref{fig:60to350}, there appears to be a
correlation between colour and cirrus RMS for the sources classified as extended
in the ERCSC. We conclude that the ERCSC fluxes for these sources are partially
contaminated with cirrus emission from our own Galaxy. We thus exclude these 127
sources from further analysis. Of the remaining 468 non-extended ERCSC sources,
fewer than 10 lie in the region of this correlation. These sources are retained
for the following analysis, but any conclusions that come solely from these
specific sources will be treated with caution.

More generally, this analysis highlights one of the issues that must be faced
when using the ERCSC catalogue. Anyone wishing to cross-match \Planck\ sources,
especially those detected at high frequencies, with sources at other wavelengths,
needs to take great care in ensuring that the ERCSC fluxes are not contaminated
by cirrus emission.

\section{Comparison to Existing Submm Data}\label{sec:Comparison}

\subsection{Galaxies detected with SCUBA}

The largest studies of cool dust in external galaxies to date have been
associated with the SCUBA Local Universe Galaxy Survey (SLUGS) and its extensions
\citep{2000MNRAS.315..115D,2001MNRAS.327..697D,2005MNRAS.364.1253V,
2010MNRAS.403..274C}. These encompass a total of about 250 objects that were
observed with SCUBA. The targets were selected on the basis of {\it IRAS\/} flux
$B$-band optical magnitude or FIR luminosity. Most of the objects were detected
only at $850\,\mu$m (i.e., not also at $450\,\mu$m), allowing, with the
{\it IRAS\/} data, only a single component ($T, \beta$) fit -- where the SED is
described as $S_{\nu} \propto \nu^{\beta} B(\nu, T)$, with $B(\nu, T)$ being the
\Planck\ function, and the parameters $T$ and $\beta$ representing temperature and
dust emissivity index, respectively. A small fraction of SLUGS galaxies were also
detected at $450\,\mu$m, allowing for the existence of a second, cooler, dust
component to be assessed. For these objects, and more recently for an
ultra-luminous IR galaxy (ULIRG) sample,
\cite{2001MNRAS.327..697D} and \cite{2010MNRAS.403..274C} found some evidence for
a colder dust contribution.

\begin{figure}
\centering
\includegraphics[angle=0,scale=0.52]{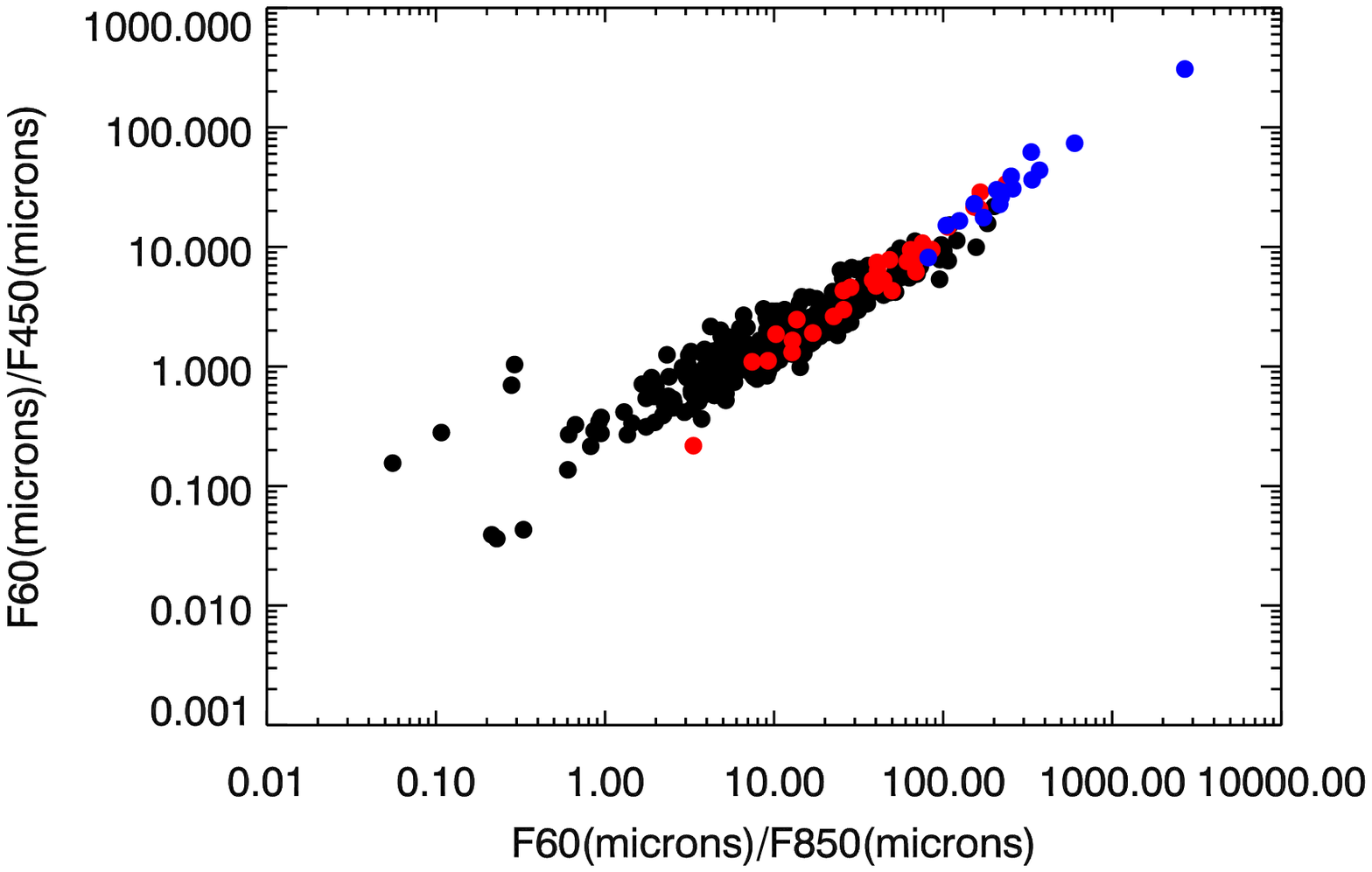}
\caption{Colours for the ERCSC galaxies (black dots) compared to those found for
SLUGS galaxies \citep[red;][]{2001MNRAS.327..697D,2005MNRAS.364.1253V}
and ULIRGs \citep[blue;][]{2010MNRAS.403..274C}. A flux correction
factor of 0.506 has been applied to the \Planck\
\getsymbol{HFI:center:frequency:857GHz}\,GHz flux densities to extrapolate them
to the SCUBA $450\,\mu$m band.  Only sources detected at $>3\sigma$ in the
\getsymbol{HFI:center:frequency:353GHz}\,GHz ($850\,\mu$m) band and at $>5\sigma$
in the \getsymbol{HFI:center:frequency:857GHz}\,GHz band (the requirement for
inclusion in our analysis) are shown. The four points above the general trend in
the lower left of the diagram are the well-known non-thermal dominated sources
3C279, 0537$-$441, OJ$+$287 and 3C273. The SLUGS point with the lowest $60\,\mu$m
to SCUBA flux ratios corresponds to the galaxy IC979; it is offset from the
general correlation for SLUGS and ERCSC galaxies, and \cite{2005MNRAS.364.1253V}
note that its {\it IRAS\/} flux densities should be treated with caution. 
}
\label{fig:colours}
\end{figure}

The presence of colder dust can be inferred from colour-colour diagrams when two
submm flux densities are available. We show the SLUGS sources and the ERCSC
sources (after colour corrections to the \Planck\ flux densities and a suitable
scaling has been applied to convert from \Planck\ 857\,GHz flux density to the
SCUBA $450\,\mu$m band) in Fig.~\ref{fig:colours}.  As can be seen, the \Planck\
galaxies lie on the same broad trend as the SLUGS galaxies (with the exception of
a small number of objects dominated by a non-thermal AGN component, such as
3C273 and 3C279). The ERCSC sources, though, extend the trend to cooler FIR/submm
colours than were found for the SLUGS objects, suggesting that the galaxies
detected in the ERCSC contain cooler dust than was detected in the majority of
SLUGS sources.

\subsection{CO Contamination}

One factor that has complicated the interpretation of ground-based submm observations of galaxies has been the presence of CO emission lines within the submm passbands that make a significant contribution to the continuum flux. \cite{2004MNRAS.349.1428S} estimated that the CO (3-2) line contributed an average 25\% of the flux received in the SCUBA $850\,\mu$m continuum passband for galaxies observed in the SLUGS survey, with the range of flux contributions going from 10--45\% for the subset of SLUGS galaxies for which CO(3-2) observations were available. The SCUBA $850\,\mu$m filter has a bandwidth of $\sim$30\,GHz. The \Planck\  \getsymbol{HFI:center:frequency:353GHz} filter is significantly broader, at $\sim$90\,GHz, so the line contribution will be correspondingly smaller at $\sim$8\% on average. This contamination fraction can be checked using observed values of integrated CO 3-2 line fluxes for a variety of objects from \cite{2006A&A...460..467B} and matching them to continuum observations of similar beamsize to the CO observations. We find contamination fractions of 2\% for Arp220, 6\% for Mrk231 and a worst case example in the central region of NGC 253, where a contamination of 11\% is calculated.

We can extend this analysis to other \Planck\ bands using NGC 253 as a worst case since, unlike most others, this object has been observed over the full range of CO transitions accessible from the ground. We find that the higher frequency bands have a reduced level of contamination compared to the  \getsymbol{HFI:center:frequency:353GHz} channel, with $<$1\% at  \getsymbol{HFI:center:frequency:857GHz} and 6\% at  \getsymbol{HFI:center:frequency:545GHz} (assuming a flat spectral line energy distribution to estimate the contribution of the CO 5-4 line that is inaccessible from the ground). More normal objects than NGC 253, not dominated by an ongoing starburst, will have an even smaller level of CO copntamination than this. At lower frequencies, though, the contamination can become more serious. CO 2-1 could contribute as much as 21\% of the continuum flux in the  \getsymbol{HFI:center:frequency:217GHz} band in the inner regions of NGC253. In the  \getsymbol{HFI:center:frequency:100GHz} band for this object the CO line could contribute as much as 75\% of the flux of the thermal continuum. NGC253 is a worst case scenario, so more typical sources would of course suffer much less contamination. However, very few galaxies are detected by \Planck\ solely in thermal emission in this band, and the few that are detected are bright nearby objects with substantial archival data that can allow a direct assessment of the CO contribution. 

Our conclusion from this analysis is that the CO contribution to the continuum flux is likely to be smaller than other sources of uncertainty for generic ERCSC-detected galaxies except for the small number which are detected in the  \getsymbol{HFI:center:frequency:217GHz} band. Flux excesses detected in this band alone might thus result from CO emission rather than from any putative very cold dust component.

\section{Parametric Models of Dust SEDs}\label{sec:Parametric}

\subsection{Fitting Method}
\label{sec:sed_fitting}

Given the \Planck\ and {\it IRAS\/} flux data described in Section~\ref{sec:Observations}, with appropriate colour corrections applied to the \Planck\ flux densities, we model the underlying signal in observed frequency band $\nu$ as one or more grey-body sources with flux density
\begin{equation}
	G(\nu; T, \beta) \propto \nu^\beta B_\nu(T)\;,
\end{equation}
where $B_\nu(T)$ is the \Planck\ function for blackbody flux density. We fit the data $d_\nu$ to one-component models of the form
\begin{equation}\label{eq:singleT}
	d_\nu = A G[\nu(1+z); T, \beta] + n_\nu\;,
\end{equation}
or to two-component models with a fixed $\beta=2$ grey-body exponent,
\begin{equation}\label{eq:twoT}
	d_\nu = A_1 G[\nu(1+z); T_1, 2] + A_2 G[\nu(1+z); T_2, 2] + n_\nu\;.
\end{equation}
In these equations, $A$ or $A_i$ is an overall amplitude for each component, and the factor of $(1+z)$ converts from rest-frame frequency to observed frequency for an object at redshift $z$. The noise contribution is given by $n_\nu$, which we model as a Gaussian with variance $\sigma^2_\nu$. For the \Planck\ channels, the determination of the noise contribution is described in \cite{planck2011-1.10}. For {\it IRAS}, the detections are classified in the IIFSCZ of Wang \& Rowan-Robinson (2009) into (1) good detections, for which we take $\sigma_\nu = 0.1 d_\nu$; (2) marginal detections, for which we take $\sigma_\nu = 0.5 d_\nu$; and (3) upper limits, for which we take $\sigma_\nu$ to be the reported upper limit, and $d_\nu=0$. As mentioned above, we only consider sources with detections in the \getsymbol{HFI:center:frequency:857GHz} and  \getsymbol{HFI:center:frequency:545GHz} bands at significance of at least 5$\sigma$ or greater and in the  \getsymbol{HFI:center:frequency:353GHz} band of at least 3$\sigma$.

Thus, the parameters of our model are some subset of the $A_i, T_i, \beta$, depending on which model we fit. We use a Bayesian Markov Chain Monte Carlo (MCMC) \citep[e.g.,][]{LewisBridle2002,Jaynes2003} technique to probe the parameter space; with our Gaussian noise, this is equivalent to an exploration of the $\chi^2$ surface, albeit with a nonlinear parameterization. We require a $0\le\beta\le3$  and $3\,{\rm K}\le T \le 100\,{\rm K}$ with a uniform prior probability between those limits (detections of very low temperatures, $T<10\,{\rm K}$, are actually dominated by non-thermal emission). We adopt a uniform prior on $\ln A_i$, as it ranges over many orders of magnitude for sources of widely varying absolute luminosities and distances.

The MCMC engine first creates a 15,000-sample Markov chain, varying one parameter at a time, using this to find an approximately-orthogonal linear combination of parameters, with which a subsequent 100,000-sample chain is run. Convergence is assessed by re-running a small number of chains from a different starting point and checking for agreement to much better than one standard deviation in all parameters.

We calculate an approximation to the Bayesian evidence, or model likelihood \citep{Jaffe1996,Jaynes2003} in order to compare the two-temperature and one-temperature fits. The evidence is calculated as the average of the likelihood function over the prior distribution; we approximate the likelihood as a multivariate Gaussian function of the parameters centred at the maximum-likelihood MCMC sample with covariance given by the empirical covariance of the samples (this approximation ignores the prior on the amplitude of the individual grey bodies). 

\begin{figure*}[htbp]
	\centering
\includegraphics[width=\columnwidth]{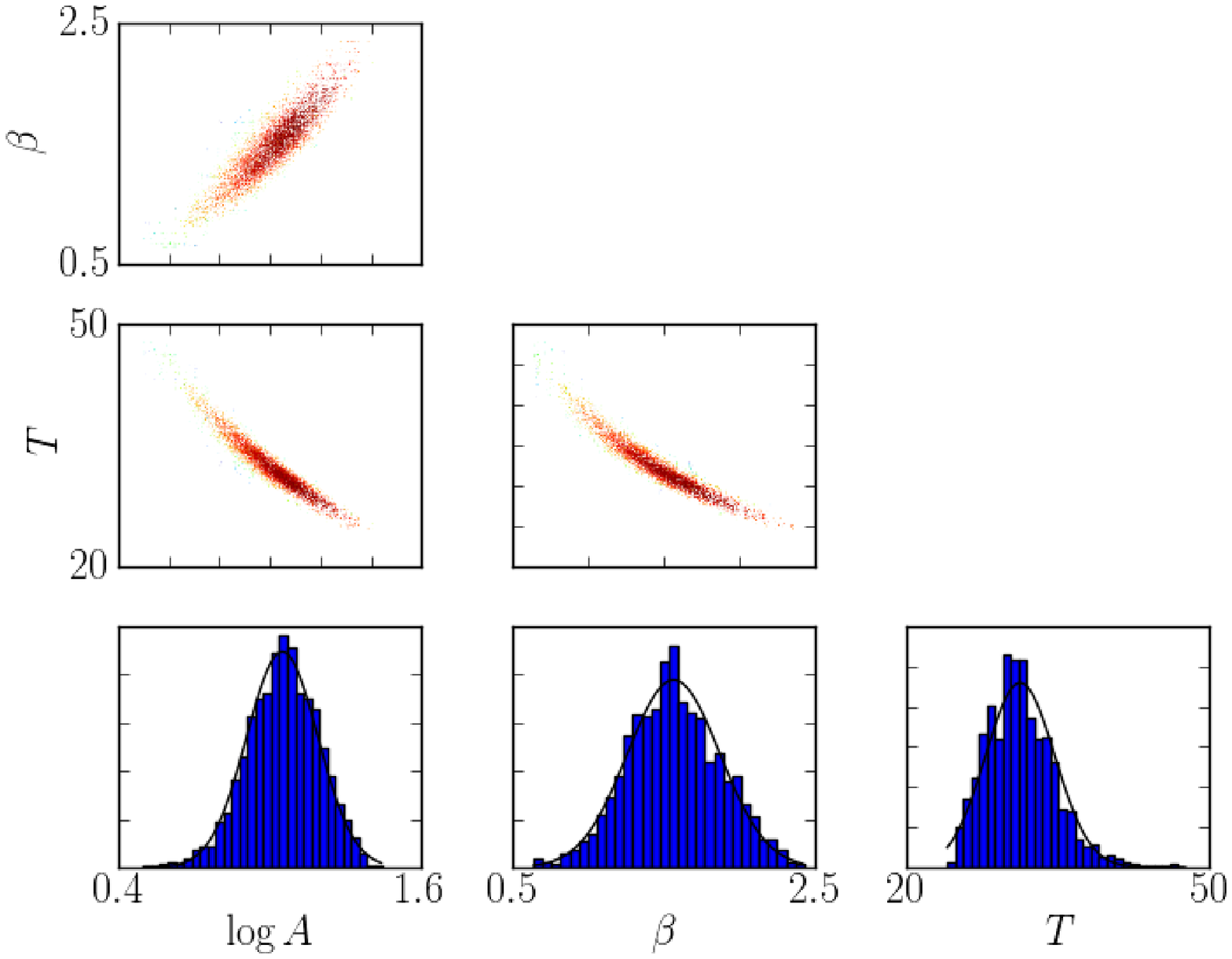}
\includegraphics[width=\columnwidth]{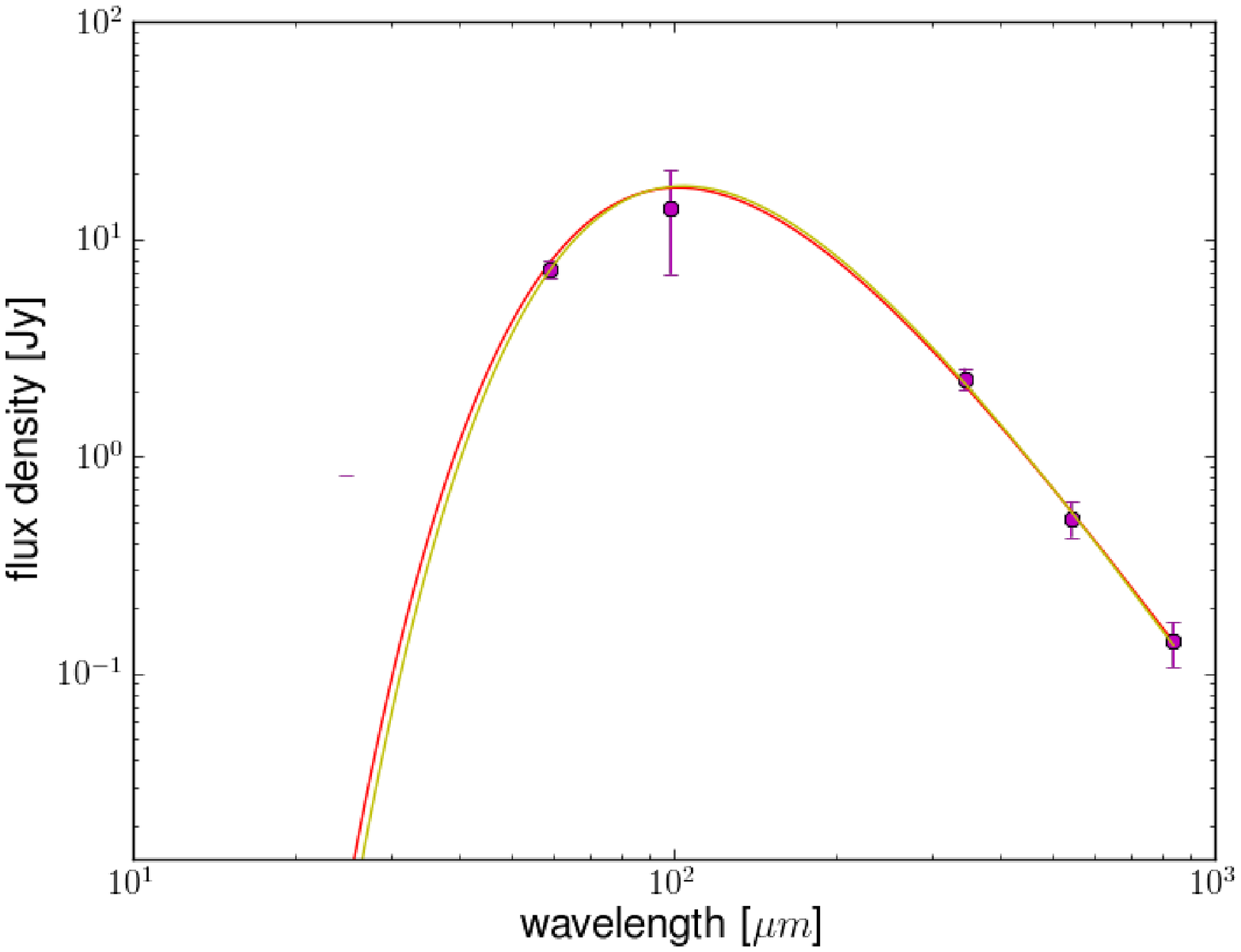}
	\caption{Left panel: Samples from the SED likelihood function ($\chi^2$) for F01384-7515 and the one-temperature model. The bottom row shows the one-dimensional (marginalized) posterior for the parameters ($\log A$, $\beta$, $T$), and the other panels show all two-dimensional marginal distributions. Right panel: the data points for this object, along with the one-temperature model for the maximum likelihood sample (yellow curve; $T=30,{\rm K}$, $\beta=1.7$) and the model determined by the mean of the samples of each parameter (red curve; $T=(32\pm5)\,$K, $\beta=1.6\pm0.4$ where the errors are the variances). Note the logarithmic axes which make interpretation of the error bars difficult.}
	\label{fig:OneT}
\end{figure*}

\begin{figure*}[htbp]
	\centering
	\includegraphics[width=\columnwidth]{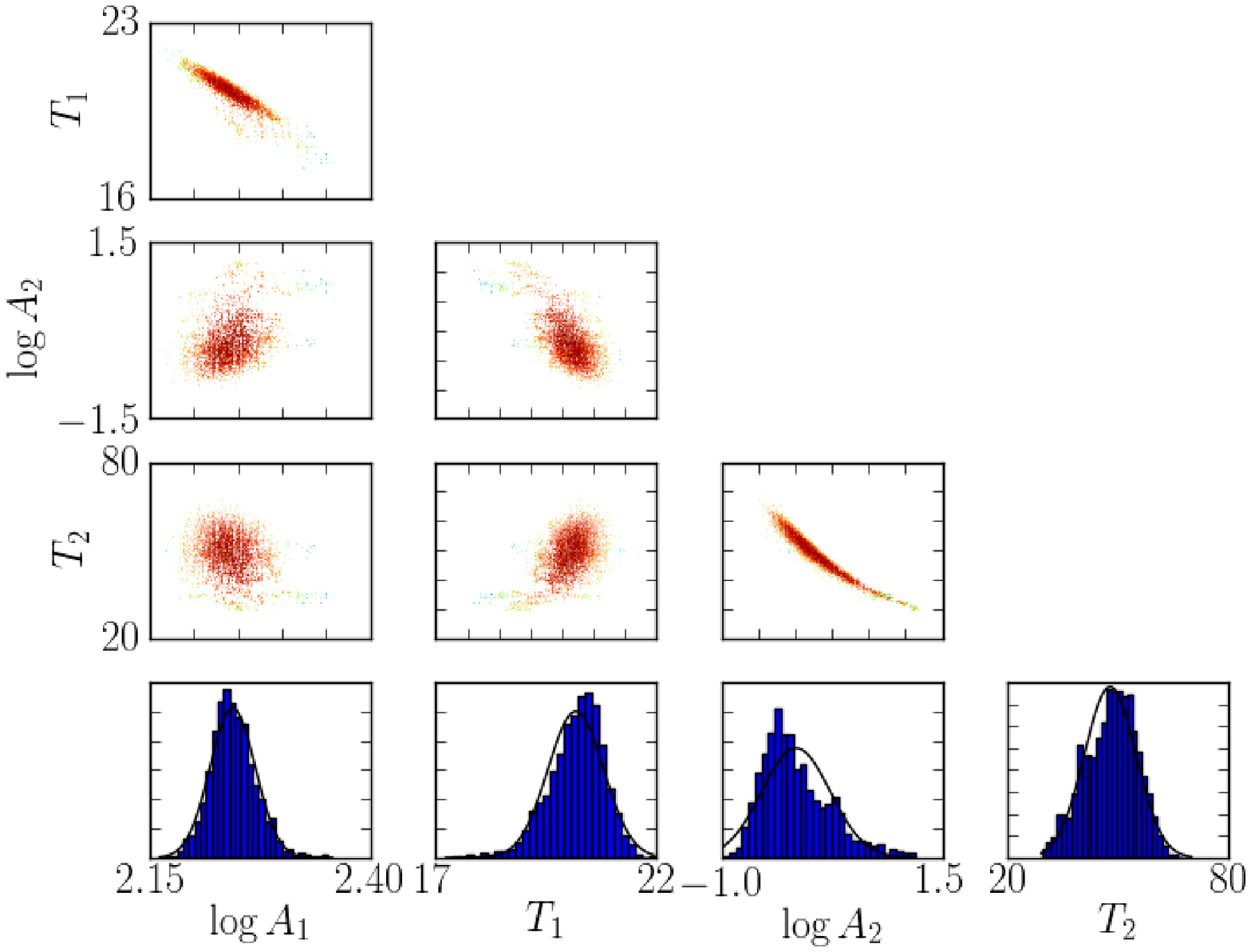}
	\includegraphics[width=\columnwidth]{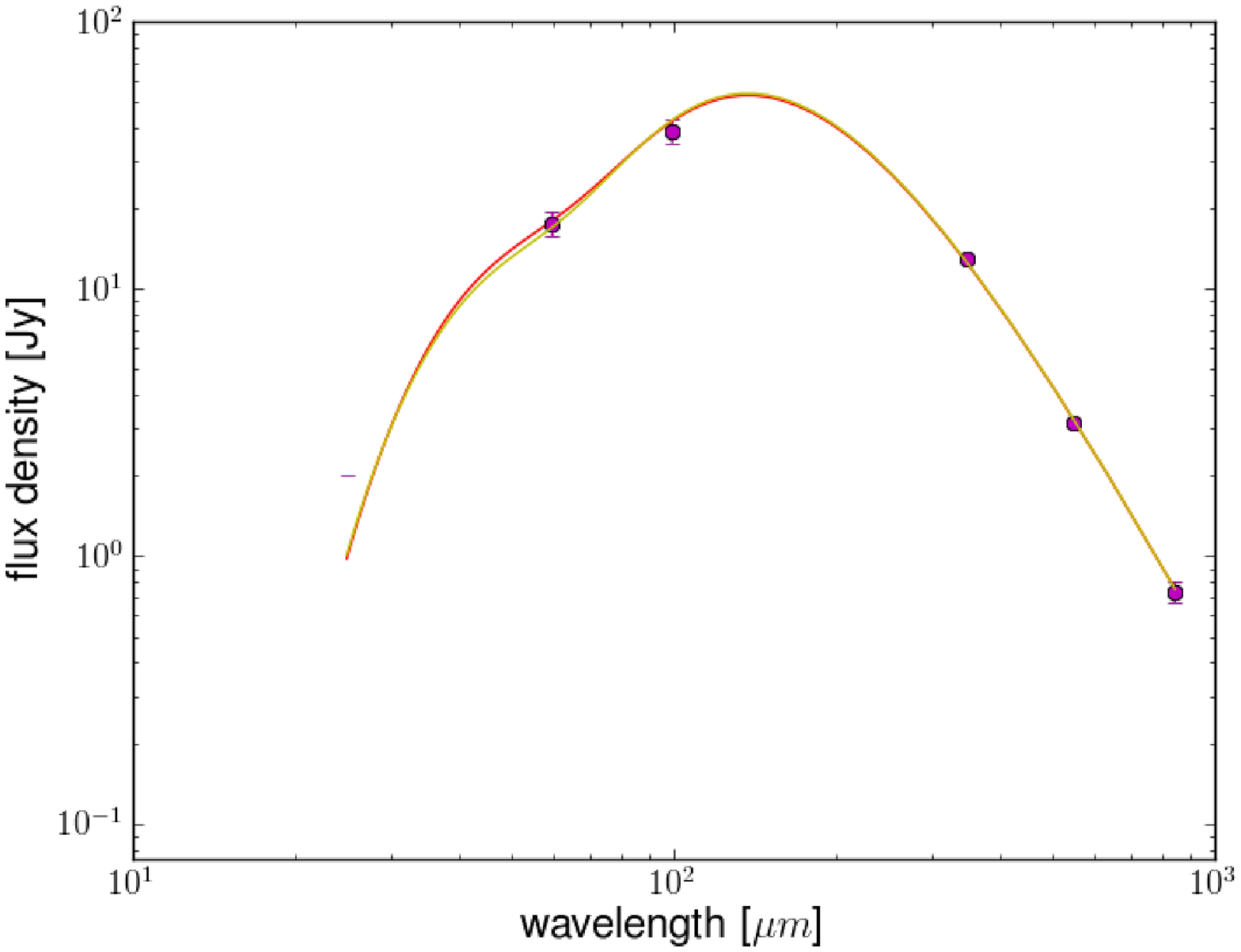}
	\caption{As in Figure~\ref{fig:OneT}, but for F00322-0840 and the two-temperature model with fixed $\beta=2$. The maximum-likelihood temperatures are $20\,$K and $49\,$K the means and variances are  $(20\pm0.8)\,$K and $(44\pm8)\,$K.}
	\label{fig:TwoT}
\end{figure*}

In Figures~\ref{fig:OneT},~\ref{fig:TwoT}  we show sample output from our MCMC runs for different objects and models, along with the measured SEDs and fits. For objects such as F01384-7515 in Figure~\ref{fig:OneT}, if we instead perform a two-temperature fit, it prefers the amplitude of the second temperature component to be many tens of orders of magnitude below the first, and gives temperature values consistent with the one-temperature fit; this indicates, along with the approximate evidence discussed above, that a one-temperature model is strongly preferred.

\subsection{Results from Parametric Fits}

\subsubsection{Single Component Fits}

Figure 7 shows the $T-\beta$ plane for parametric fits to all 468 reliably identified non-extended sources within the ERCSC-IIFSCz cross match whose flux densities
pass our S/N ratio criteria for inclusion, together with similar single temperature parametric fits from  \cite{2000MNRAS.315..115D} and \cite{2010MNRAS.403..274C}. As can be seen, the \Planck\ ERCSC sources overlap with the SLUGS galaxies but extend to cooler temperatures and flatter, i.e., lower $\beta$, SEDs. The median parameters for the \Planck\ sources are $T = 26.3\,$K with temperatures ranging from 15 to $50\,$K, and $\beta$=1.2 compared to corresponding values from 104 SLUGS galaxies \cite{2000MNRAS.315..115D} of $T=35\,$K and $\beta$=1.3 and of $T=41\,$K and $\beta$=1.6 for 26 ULIRGs  \cite{2010MNRAS.403..274C}. This confirms the result from consideration of \Planck-{\it IRAS\/} colours in Fig.~4 that we are seeing cooler dust in the ERCSC-IIFSCz galaxies.

There are ten sources common to the ERCSC-IIFSCz cross-matched catalogue and the SLUGS studies. The fits using \Planck\ data and using SLUGS data for all but two of these sources are in good agreement. The two exceptions are NGC7541 and NGC5676. The likely cause of the disagreements in these cases, is the presence of a close companion {\it IRAS\/} source to NGC7541 (NGC7537 $3.1^\prime$ separation, and thus only $\sim$0.7 beam FWHM away), so that the \Planck\ flux density is likely over-estimated, and extended {\it IRAS\/} emission in NGC5676, making the {\it IRAS\/} FSC flux densities used in our analysis underestimates.

\begin{figure}
\centering
\includegraphics[angle=90,scale=.4]{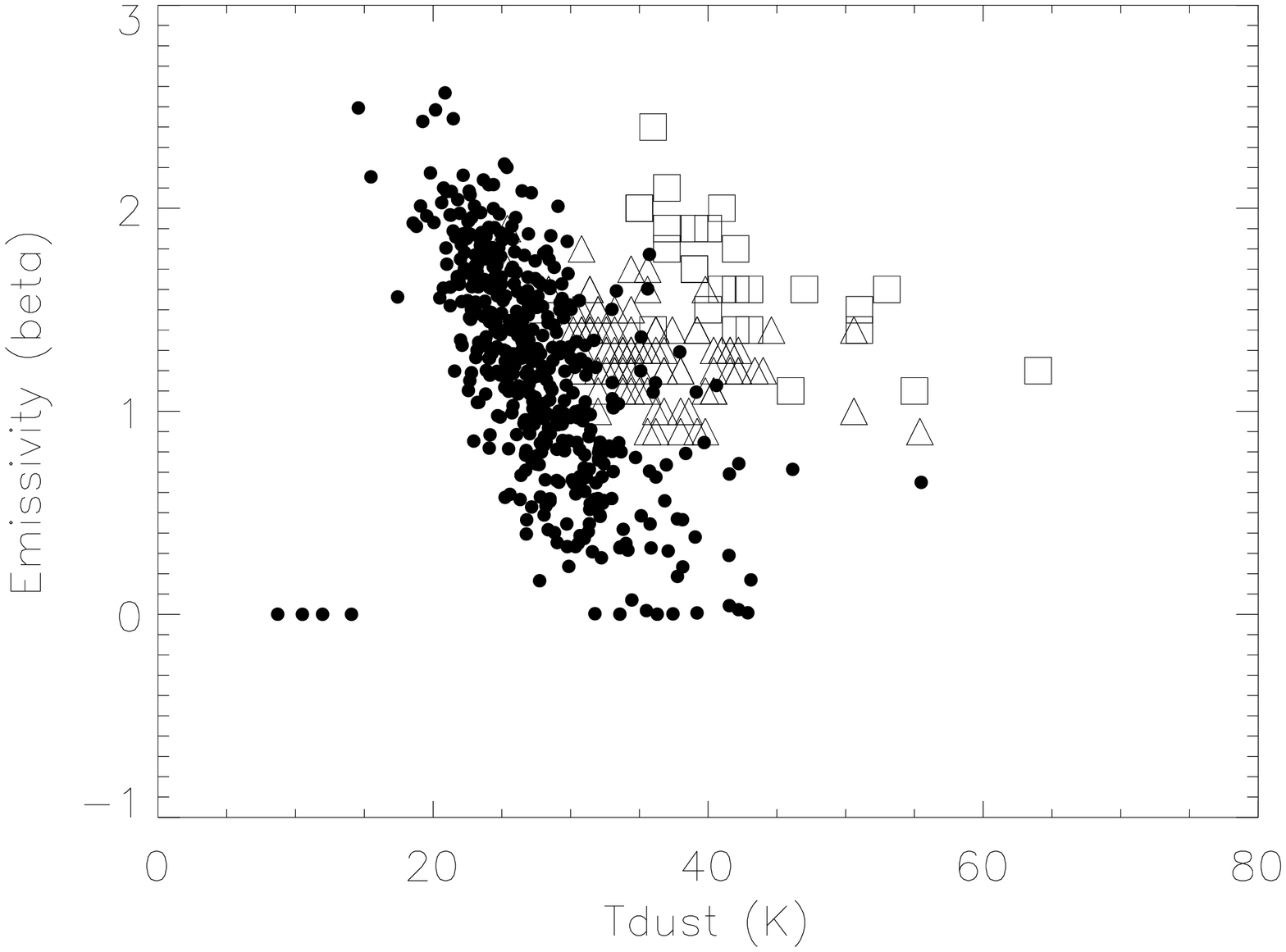}
\caption{Temperature-$\beta$ correlation for ERCSC-IIFSCz matched galaxies (solid dots) together with data from \cite{2000MNRAS.315..115D} and \cite{2010MNRAS.403..274C} for SCUBA observed sources (triangles and squares respectively). The four sources with $\beta$=0 and $T<10\,$K at the left of the diagram are the non-thermal dominated sources 3C279, 0537-441, OJ+287 and 3C273.}
\end{figure}

The position of our galaxies in the Luminosity-Temperature plane is an important question since it relates to claims of evolution in the dust properties of galaxies. It has previously been suggested that high redshift, high luminosity SMGs have lower dust temperatures and higher dust masses than more nearby objects \cite{2007ApJ...660.1198Y}. Comparison of SMGs from \cite{2005ApJ...622..772C}, \cite{2008MNRAS.384.1597C} and \cite{2006ApJ...650..592K} with more local galaxies from  \cite{2000MNRAS.315..115D}  and local ULIRGs \cite{2010MNRAS.403..274C} confirms this effect. Claims have been made that sources selected at longer wavelengths than the $60\,\mu$m typical of {\it IRAS\/} derived samples \citep[e.g.,][]{2009MNRAS.397.1728S}, Patel et al. (in prep)) show less of a separation between the local sources and the higher redshift SMGs. Much of this work is hampered by the poor sampling of the dust SEDs of local objects at wavelengths between 100 and $850\,\mu$m. Recent results from {\it Herschel\/} \cite{2010A&A...518L...9A} and BLAST \cite{2009ApJ...703..285D} have begun to fill the gap between local {\it IRAS\/} galaxies and the SMGs, suggesting that our view of dust temperatures in local objects are biased to warmer temperatures through our dependence on {\it IRAS\/} flux densities. In Fig.~8 we show the positions of \Planck\ galaxies on the $L-T$ plane. As can be seen, the gap between the local galaxies and the SMGs is starting to be filled by the \Planck\ objects. The \Planck\ fluxes for one of these sources may include some contamination from cirrus, but the rest lie in areas of normal to low Galactic cirrus noise and should thus be fully reliable. The large area coverage of \Planck\ is particularly important in this as it allows us to probe generic L/ULIRG-class objects ($L_{\rm FIR} > 10^{11}{\rm L}_{\odot}$) rather than having to rely on pre-selected {\it IRAS\/} bright sources as in \cite{2010MNRAS.403..274C}. {\it Herschel\/} observations, which are also beginning to show the gap being filled, do not cover enough area to include many such L/ULRG objects in the local Universe \cite{2010A&A...518L...9A}. We find several cool ($T<30\,$K) ULIRGs that have very similar characteristics to SMGs. The issue of the apparent distinction between local galaxies and high-$z$ SMG population thus seems to be approaching resolution.

\begin{figure}
\centering
\includegraphics[angle=90,scale=0.4]{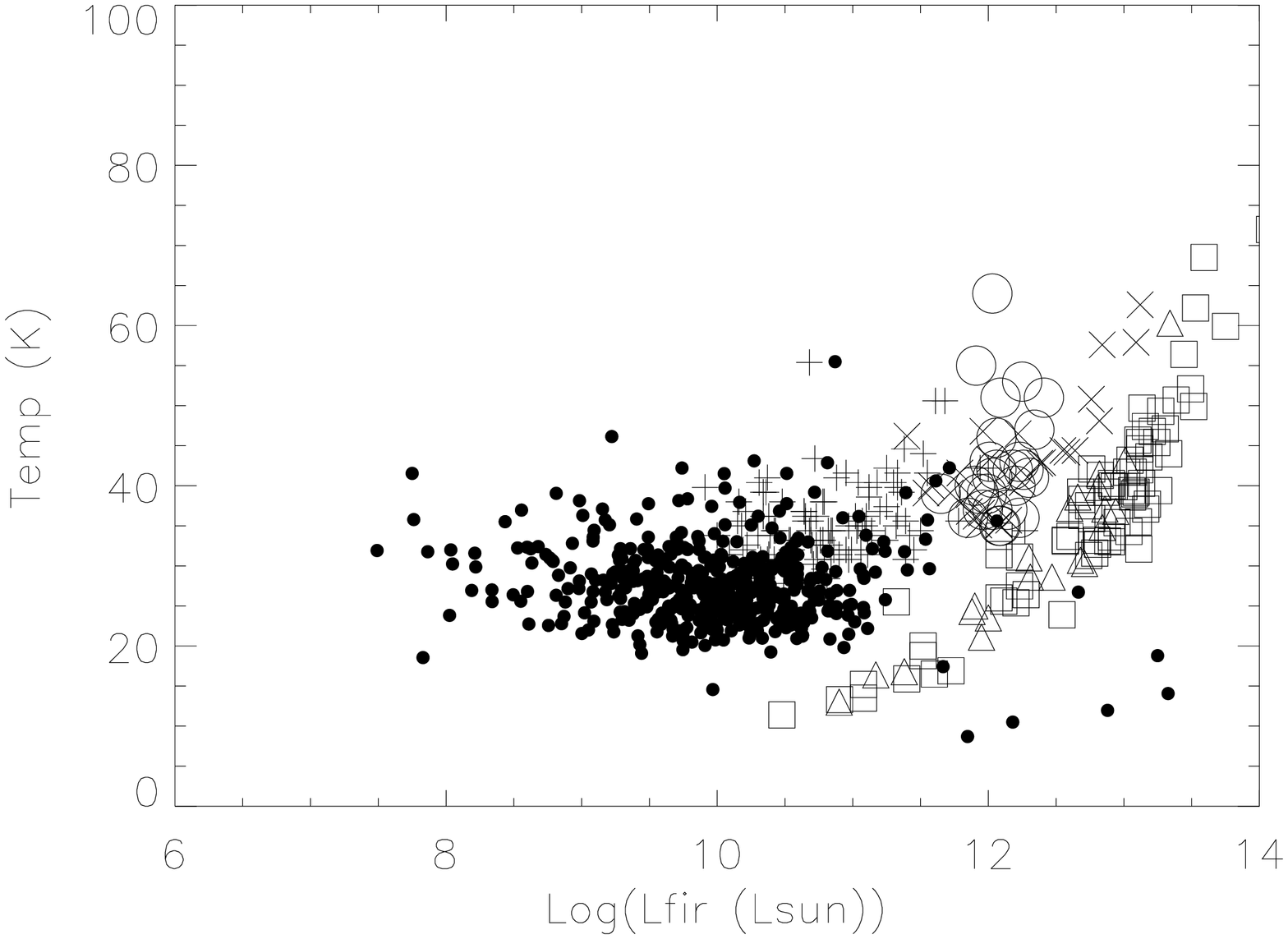}
\caption{The Temperature-Luminosity plane showing a variety of FIR populations. Open squares are SMGs from \cite{2005ApJ...622..772C}, open triangles are SMGs observed with SHARCII by \cite{2008MNRAS.384.1597C}  and \cite{2006ApJ...650..592K} . + signs are the SLUGS sources from \cite{2000MNRAS.315..115D}, x are intermediate redshift ULIRGs from \cite{2007ApJ...660.1198Y}, ULIRGs from \cite{2010MNRAS.403..274C} are open circles. \Planck-ERCSC-IIFSCz galaxies are shown as solid dots. As can be seen the previous apparent distinction between the local FIR populations and the SMGs is weakened by sources from this work lying in the same region as the SMGs and by filling in some of the gap between the populations. The four aberrant sources with $T<$15K and high $L$ in the bottom right of the plot are the non-thermal dominated sources 3C279, 0537-441, OJ+287 and 3C273.}
\end{figure}

While the temperature distribution of our objects is consistent with what has been seen elsewhere, we find that some of our galaxies have $\beta$ more than 3$\sigma$ less than 1. One possible cause for this might be the  \getsymbol{HFI:center:frequency:353GHz} flux density being affected by Eddington bias (see e.g., \cite{2004A&A...424...73T}), leading to 'flux-boosting' of lower significance detections, since we accept these fluxes down to 3$\sigma$. This is tested by repeating the fits using only the {\it IRAS},  \getsymbol{HFI:center:frequency:857GHz} and  \getsymbol{HFI:center:frequency:545GHz}  flux densities. While there are small differences in fits to individual objects resulting form the exclusion of the  \getsymbol{HFI:center:frequency:353GHz} flux densities, the general distribution remains the same, complete with the low $\beta$ sources. We thus conclude that this is a real effect and not due to Eddington bias or any other issue related to the  \getsymbol{HFI:center:frequency:353GHz} flux densities, such as contamination by emission from the CO 3-2 molecular line. While such low $\beta$ values are not expected in simple models of dust, it is suggestive that the SEDs can be better fit by a parameterization that uses a mixture of dust at two temperatures, as suggested by \cite{2001MNRAS.327..697D} and which is a good fit to our own galaxy as seen by {\it COBE\/} \cite{1995ApJ...451..188R}. We investigate this by applying two component fits to the dust SEDs.

\subsubsection{Two Temperature Fits}

We carry out two temperature component fits  on our sources, assuming $\beta$=2 for both components, and present the temperature-temperature plot in Fig.~9. We also use the Baysian evidence calculated during the fitting process \citep{Jaffe1996,Jaynes2003} to determine how many of our sources show evidence for a two component fit above that of the single component ($T,\beta$) fit. We find that the two component fit is favoured in most cases, with 425 objects giving a higher evidence for this model and only 43 preferring the single component fit. Once again we test the possibility that issues with the lower significance  \getsymbol{HFI:center:frequency:353GHz}\,GHz flux densities might bias these fits by repeating the analysis with these fluxes excluded. While this increases the uncertainties in the fits, as with the ($T,\beta$) fits we find that the exclusion of the  \getsymbol{HFI:center:frequency:353GHz}\,GHz fluxes makes no systematic difference to the temperatures found.

We find 17 galaxies fit by models containing a dust component with temperatures as low as $10\,$K. Four of these are the bright blazars and are thus dominated by non-thermal emission, while one is a source that might still contain some cirrus contamination. We thus find at least 13 galaxies which appear to contain very cold dust. Such dust has previously been found in our own galaxy \citep[e.g.][]{1995ApJ...451..188R} but has not been seen before in large scale extragalactic surveys. The small number of sources which have $T_{\rm cold}$ lower than $7\,$K are all dominated by nonthermal emission. The details of the temperature-temperature plot in Fig.~9, with small scatter and the  temperature of the hot component being largely independent of that of the cold component up to $T_{\rm cold} \sim 18\,$K will likely have implications concerning the relationship between hot and cold dust components.

\begin{figure}
\centering
\includegraphics[angle=90,scale=0.4]{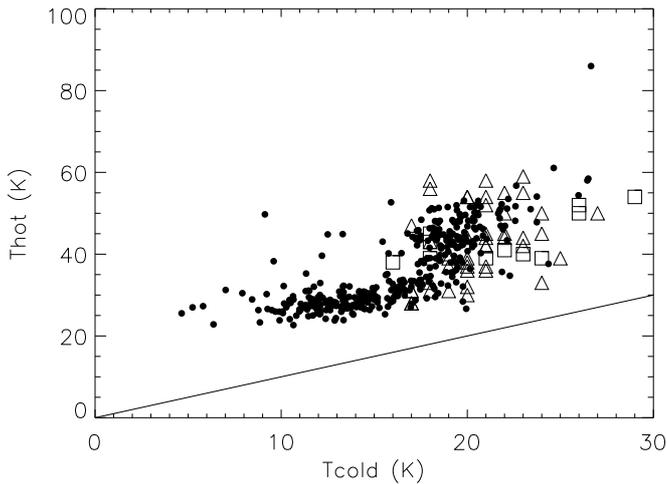}
\caption{The Temperature-Temperature plane for two temperature component fits for ERCSC-IIFSCz matched galaxies. together with data from \cite{2001MNRAS.327..697D}  and \cite{2010MNRAS.403..274C} for SCUBA observed sources (triangles and squares respectively). Only those ERCSC-IIFSCz sources where the two temperature model is preferred and with a reliably determined cold dust component temperature (i.e., $T/\sigma(T) > 4$) are plotted. The sources with the coldest $T_{\rm cold}<7\,$K in this plot are dominated by non-thermal emission.}
\end{figure}

\subsection{The Broader ERCSC-IIFSCz Sample}

The parametric fitting reported here only concerns the 468 non-extended ERCSC-IIFSCz matched galaxies which are detected at 5$\sigma$ or greater in the  \getsymbol{HFI:center:frequency:857GHz} and  \getsymbol{HFI:center:frequency:545GHz} bands and at 3$\sigma$ or greater at  \getsymbol{HFI:center:frequency:353GHz}. This ensures that uncertainties in the fluxes do not preclude a good fit to the SED, but brings the risk that we might be missing a significantly different subclass of object in the remaining 1122. We have thus applied our fitting methods to this whole sample, regardless of ERCSC S/N beyond the basic detection requirement of a 5$\sigma$ or greater detection at \getsymbol{HFI:center:frequency:857GHz}\,GHz. While there are larger errors bars on the fitted parameters we find no indication that galaxies in this larger sample have a different range of dust properties to those discussed above.

\section{Physical Models: Template Fitting}\label{sec:Physical}

The FIR and submm spectral energy distributions of galaxies from the {\it IRAS}, {\it ISO\/} and {\it Spitzer\/} surveys have been successfully modelled with a small number of templates (\cite{1992MNRAS.258..787R}, \cite{2005AJ....129.1183R}, \cite{2008MNRAS.386..697R}).  However the submm data available in such studies is quite limited and we expect to get a much better understanding of cool dust in galaxies with the data from \Planck.  Already with {\it Herschel}, there is evidence for unexpected quantities of cold dust in some galaxies \cite{2010MNRAS.409....2R}.

The \cite{2010MNRAS.409....2R}. study of the SEDs of {\it Herschel}-detected galaxies used SPIRE flux densities extending from 250 to $500\,\mu$m combined with pre-existing data from SWIRE at {\it Spitzer\/} and optical wavelengths. Through the combination of \Planck\ data with the IIFSCz, the galaxy sample considered here includes data from the optical to {\it IRAS\/} fluxes, and then the \Planck\ data has flux densities at 350, 550 and $850\,\mu$m (\getsymbol{HFI:center:frequency:857GHz},  \getsymbol{HFI:center:frequency:545GHz} and  \getsymbol{HFI:center:frequency:353GHz} GHz). Some of our objects are also detected at 1.4 mm (\getsymbol{HFI:center:frequency:217GHz} GHz) which is included in our analysis if available. The range of wavelengths available with the ERCSC-IIFSCz sample is thus broader than that available through {\it Herschel}. This enables us to place better constraints on the role and importance of cold dust in these objects. Our sample is also much larger than the 68 objects considered in \cite{2010MNRAS.409....2R}, so we can better determine the variety and overall statistics of the SEDs of local galaxies.

The templates used in fitting {\it IRAS}, {\it ISO\/} and {\it Spitzer\/}
data are (1) a ÔcirrusÕ (optically thin interstellar dust) model characterised by a radiation intensity
$\phi = I(galaxy)/I(ISRF) = 5$, where I(ISRF) is the intensity of the radiation field in the solar neighbourhood,
(2) a normal M82-like starburst, (3) a higher optical depth Arp220-like starburst, (4) an AGN dust torus.
In fitting the SEDs of {\it Herschel\/} galaxies,  \cite{2010MNRAS.409....2R} use two further cirrus templates with $\phi$ = 1 and 0.1, which correspond to significantly cooler dust than in the standard cirrus template.  The starting point for the present analysis is the template fit for each object given in the IIFSCz Catalogue \cite{2009MNRAS.398..109W}. This was done by fitting the optical and near-IR fluxes with an optical galaxy or QSO template.  The {\it IRAS\/} data were then fitted with one of the original four \cite{2008MNRAS.386..697R} templates. This model is then compared to the \Planck\ data. In almost all cases additional components are needed, since the {\it IRAS}-based predictions underestimate the submm flux densities provided by \Planck. More weight is given to  \getsymbol{HFI:center:frequency:857GHz} and  \getsymbol{HFI:center:frequency:545GHz} flux densities in this process since they are generally at higher signal to noise and are thus less subject to Eddington bias effects. The  \getsymbol{HFI:center:frequency:353GHz}GHz and, where detected,  \getsymbol{HFI:center:frequency:217GHz}GHz flux densities are also subject to contamination by CO 3-2 and 2-1emission, respectively, which may also contribute to an excess flux in these bands beyond what might be expected from the continuum fit.

\subsection{Results from Template Fits}

In Figure 9 we analyze the SEDs of the archetypal nearby galaxies, M51, M100, M2 and Arp 220. 
M51 and M100 are modelled with two cirrus templates with $\phi$  = 1 (solar neighbourhood) and 5 (Galactic Centre), and with a modest ÔM82Õ starburst component.  M82 itself needs an additional component of cool cirrus ($\phi$ = 1) as well as the ÔM82Ó template of Efstathiou and Rowan-Robinson.  Finally Arp220 is modelled extremely well over all infrared and submm wavelengths by the ÔArp220Õ template used by Efstathiou and Rowan-Robinson.  The models for these galaxies by \cite{1998ApJ...509..103S} are also shown and perform well, especially for M51 and Arp220.

\begin{figure*}
\centering
\includegraphics[angle=0,scale=0.75]{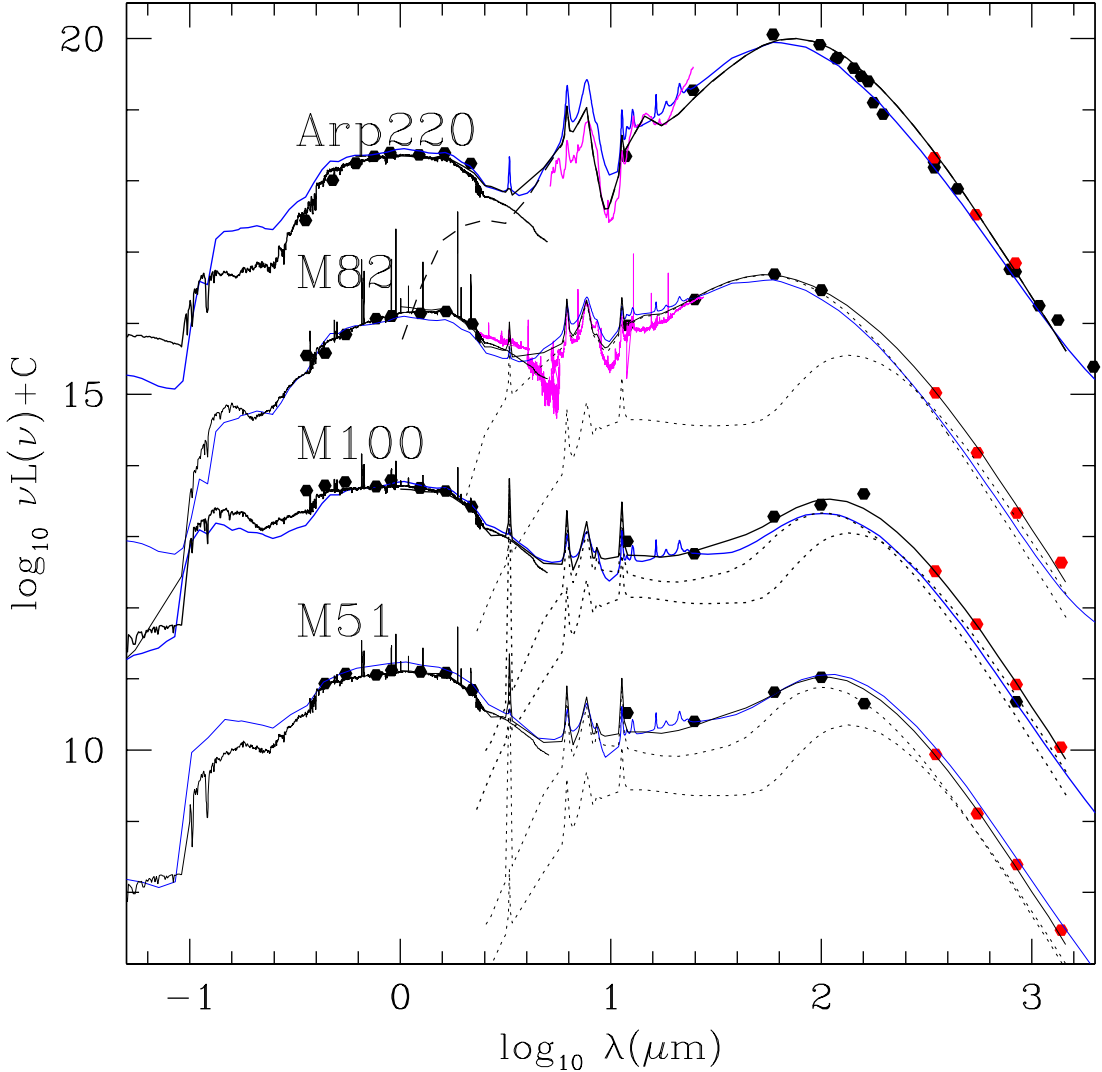}
\caption{
Template fits for the four archetypal nearby galaxies, M51, M100, 
M82 and Arp 220.  Black curves: fits with Efstathiou and Rowan-Robinson 
templates (black, separate components as dotted lines), blue curves:  \cite{1998ApJ...509..103S} models. \Planck\ ERCSC 
data shown as red filled hexagrams.  {\it ISO}-SWS 
mid-infrared spectroscopy data for M82 and {\it Spitzer}-IRS data for Arp 220 
\citep{2007A&A...461..445S} are shown in magenta.
}
\end{figure*}

Fig.~10 shows fits to galaxies with detections in 16 photometric bands: 5 optical
bands (SDSS), 3 near infrared bands (2MASS), 4 mid and FIR bands
({\it IRAS\/}) and 4 submm bands (\Planck).   The blue curve is the solar
neighbourhood cirrus template ($\phi$ = 1) and contributes significantly to the
SEDs of 7 out of the 8 galaxies.  Dust grains in this component are in the range
15--20\,K, depending on grain radius and type (Rowan-Robinson 1992).  Dust masses
are in the range $10^7 - 3.10^8 {\rm M}_{\odot}$.

\begin{figure*}
\centering
\includegraphics[angle=0,scale=0.75]{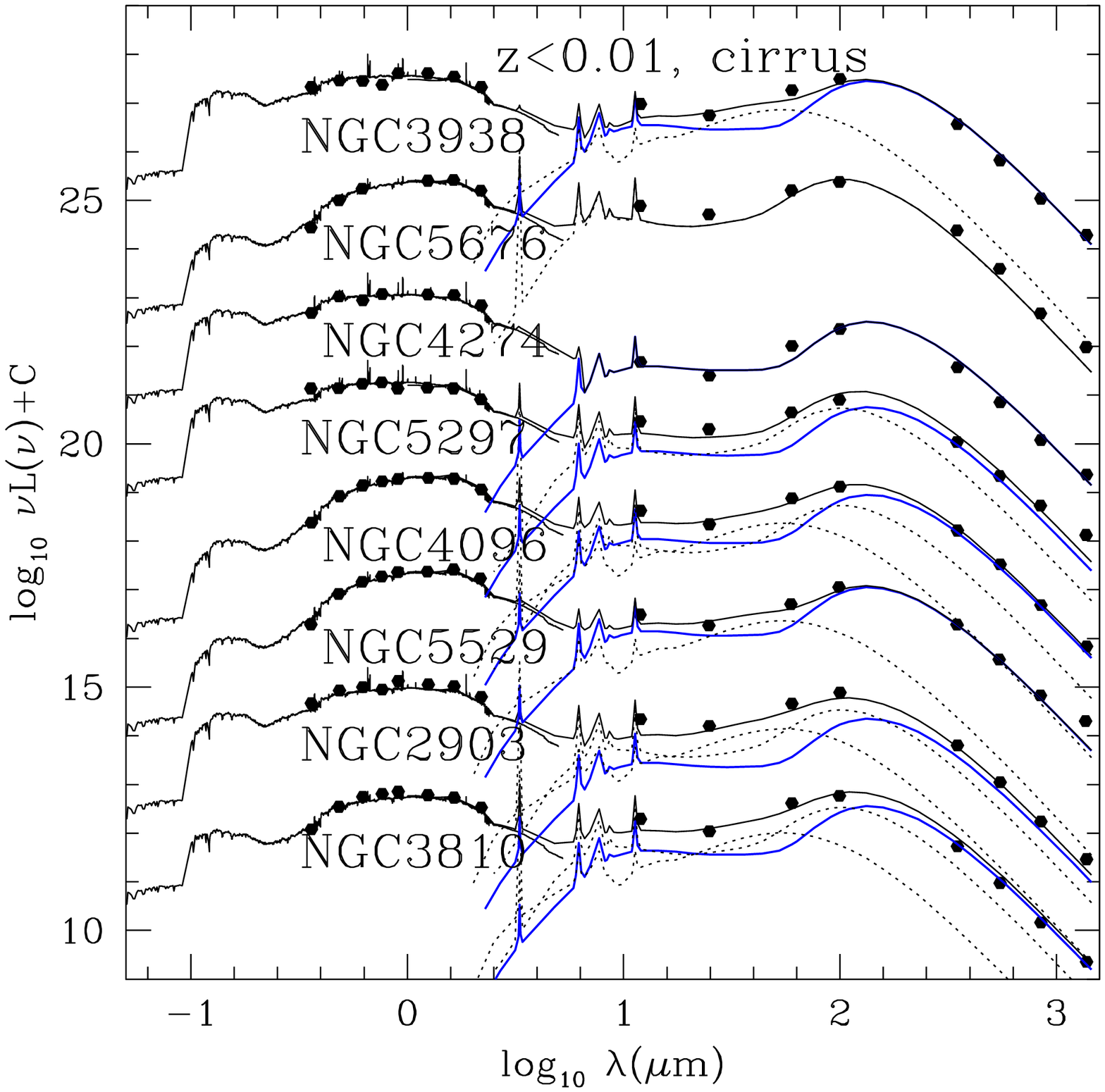}
\caption{Template fits for 8 nearby galaxies with detections in 16 bands.  Blue curve is cirrus template used for solar neighbourhood ($T_{\rm dust}=15$--$20\,$K, $\phi$ = 1), other components shown as dotted lines.}
\end{figure*}

To get an overview of the whole sample, Fig.~12 shows the distribution of  \getsymbol{HFI:center:frequency:545GHz} GHz ($550\,\mu$m) flux density versus redshift for galaxies well detected at 350--$850\,\mu$m, and with spectroscopic redshifts.  The loci of galaxies with an Arp 220 template at luminosity $L_{\rm IR} =10^{12}$ and $10^{13} {\rm L}_{\odot}$ are shown.  With the restrictions to point-sources with good detections
in the three highest frequency  \Planck\ bands, and to galaxies with spectroscopic redshifts, a number of ULIRGs are found in
the ERCSC survey, but no HLIRGs, apart from the quasar 3C273.

Fig.~13 shows the colour-ratio $S_{\getsymbol{HFI:center:frequency:857GHz}}/S_{\getsymbol{HFI:center:frequency:545GHz}}$ versus redshift.  Galaxies in the {\it IRAS\/} Bright Galaxy Sample are indicated as red dots.  Sources for which $T_2<10\,$K in the 2-temperature fits discussed above are indicated as blue dots.  Sources with $log_{10} (S_{\getsymbol{HFI:center:frequency:857GHz}}/S_{ \getsymbol{HFI:center:frequency:545GHz}}) < 0.4$ represent a new population of cooler submm sources. 
We have modelled all 17 galaxies with cool  \getsymbol{HFI:center:frequency:857GHz} GHz$/$\getsymbol{HFI:center:frequency:545GHz} GHz colours ($log_{10} (S_{\getsymbol{HFI:center:frequency:857GHz}}/S_{\getsymbol{HFI:center:frequency:545GHz}} < 0.4$), and good optical
data (Fig.~14).   Almost all require the very cold cirrus model with  $T_{\rm dust} =10$--$13\,$K, $\phi = 0.1$ (green curves in Fig.~14).
Table 1 gives the properties of the galaxies whose SEDs we have modelled in detail.  The columns are:
ERCSC RA and Dec, {\it IRAS\/} FSC name, separation between ERCSC and
IIFSCz position, spectroscopic redshift, luminosity in cirrus $\phi=5$, $\phi=1$ and $\phi=0.1$ components,
luminosity in M82 and Arp220 starburst components, optical luminosity, optical SED type, extinction, stellar mass and
dust mass.  Stellar mass and dust mass are calculated as in \cite{2008MNRAS.386..697R}, but there may be some model-dependencies in the dust masses since we have assumed that the cold dust in these galaxies is similar to Galactic dust.
We can summarize the SED modelling shown in Figs 10, 11 and 14 as follows: (1) most nearby galaxies show evidence for dust at temperatures similar to that seen in the solar neighbourhood ($\phi = 1$), as well as the warmer dust found in {\it IRAS\/}, {\it ISO\/} and {\it Spitzer\/}
studies; (2) there is a new population of cool submm galaxies with even cooler dust ($\phi=0.1, T_{dust} = 10$--$13\,$K).
This cooler dust is likely to have a more extended spatial distribution then generally assumed for the gas and dust in galaxies.

\begin{table*}
\caption{Parameters for \Planck\ ERCSC-IIFSCz galaxies with SED fits}
\begin{tiny}
\begin{tabular}{rrrrrrrrrrrrrrr}\hline \hline
\\
RA& Dec & FSS & dist & ${\rm z_{spect}}$ & log(L) & log(L)& log(L)  & log(L) & log(L) & log(L)& type & {\rm $A_V$} & log(M$_*$) & log(M$_{\rm dust}$)\\
&&&&& cirr& cirr& cirr& SB & Arp220 & Opt&&&${\rm M}_{\odot}$ & ${\rm M}_{\odot}$\\
&&&&&$\psi=5$&$\psi=1$&$\psi=0.1$\\
\\ \hline \\
\multicolumn{15}{c}{Template}\\ \hline
\\~\\
202.46930 & 47.19290 & F13277+4727& 0.1 & 0.001544 & 10.05 & 9.49 & & 9.60 & & 10.35 & Sbc & 0.0 & 10.75 & 7.54\\
&&M51\\
185.72728 & 15.82777  & F12203+1605& 0.4 & 0.005240 & 10.53 & 10.19 & & 9.98 & & 11.22 & Sbc & 0.0 & 11.62 & 8.15\\
&&M100\\
148.97540 & 69.68221& F09517+6954& 0.2 & 0.000677 & & 9.19 & & 10.41 & & 9.95 & E & 0.0 & 10.54 & 7.18\\
&&M82\\
233.74080 & 23.49795 & F15327+2340& 0.4 & 0.01813   & & & & & 12.15 & 10.58 & E & 0.0 & 11.17 & 8.15\\
&&Arp220\\
\\ \hline \\
\multicolumn{15}{c}{16-band, z $<$0.01}\\ \hline
\\~\\
175.24310 & 11.47154  & F11383+1144& 0.1 & 0.003312 & 9.70 & 9.69 & & 9.24 & & 10.00 & Sab & 0.0 & 10.57 & 7.58\\
&& NGC3810 \\
143.03957 & 21.50516 & F09293+2143& 0.2 & 0.001855 & 9.70 & 9.49 & & 9.38 & & 10.41 & Scd & 0.2 & 10.79 & 7.43\\
&& NGC2903 \\
213.89354 & 36.22261 & F14134+3627& 0.3 & 0.009591& & 10.19 & & 9.60 & & 10.70 & Sab & 0.65 & 11.27 & 8.44\\
&& NGC5529 \\
181.50023 & 47.47356 & F12034+4745& 0.3 & 0.001888 & 8.90 & 9.09 & & 8.60 & & 9.65 & Sab & 0.6 & 10.22 & 6.95\\
&& NGC4096\\
206.59677 & 43.86664 & F13443+4407& 0.5 & 0.008036 & 9.90 & 9.99 & & & & 10.60 & Scd & 0.4 & 11.04 & 7.86\\
&& NGC5297 \\
184.95674 & 29.61180  & F12173+2953& 0.7 & 0.003102 & & 9.64 & & & & 10.40 & Scd & 0.0 & 10.94 & 7.45\\
&& NGC4274 \\
218.18851 & 49.45234 & F14310+4940& 0.4 & 0.007052 & 10.60 & & & & & 10.70 & Sab & 0.4 & 11.27 & 7.71\\
&& NGC5676 \\
178.20813 & 44.12143  & F11502+4423& 0.5 & 0.002699 & & 9.59 & & 9.10 & & 9.90 & Scd & 0.5 & 10.34& 7.40\\
&& NGC3938\\
\\ \hline \\
\multicolumn{15}{c}{Cold log${(S_{857}/S_{545})}<0.4$}\\ \hline
\\~\\
159.78535 & 41.69160 & F10361+4155 & 2.0 & 0.002228 & & & 8.52 & 8.10 & & 7.50 & sb & 0.2 & 7.52 & 7.33 \\
179.31494 & 49.28748 & F11547+4933 & 0.4 & 0.002592 & & 8.79 & & & & 8.70 & sb & 0.0 & 8.72 & 6.60 \\
218.19263 & 9.88910  & F14302+1006 & 0.6 & 0.004574 & 9.40 & & 8.92 & & & 9.60 & Scd & 0.2 & 10.04 & 7.76 \\
208.22507 & -1.12087 & F13503-0052 & 0.4 & 0.004623 & & 9.64 & & & & 9.70 & Scd & 0.5 & 10.14 & 7.45\\
148.79717 & 9.27136 & F09521+0930 & 0.0 & 0.004854 & & & 8.92 & 9.40 & & 9.60 & Scd & 0.0 & 10.04 & 7.73\\
219.79150 & 5.25526 & F14366+0534 & 0.4 & 0.005020 & & 9.09 & 8.62 & 8.70 & & 10.35 & Scd & 0.2 & 10.79 & 7.54\\
158.12950 & 65.03790 & F10290+6517 & 0.4 & 0.005624 & 9.40 & & 8.62 & & & 9.80 & Scd & 0.2 & 10.24 & 7.48\\
210.53130 & 55.79348 & F14004+5603 & 1.4 & 0.006014 & 9.05 & & 8.67 & & & 10.25 & Scd & 0.7 & 10.69 & 7.50\\
208.72266 & 41.30995 & F13527+4133 & 0.0 & 0.007255 & 9.51 & 8.40 & & & & 9.96 & Scd & 0.2 & 10.40 & 6.76\\
205.57553 & 60.77630 & F13405+6101 & 0.3 & 0.007322 & & 9.20 & 8.63 & & & 9.61 & Sbc & 0.5 & 10.01 & 7.58\\

228.37207 & 58.49204  & F15122+5841 & 1.0 & 0.008474 & 9.26& & 8.95 & & & 9.51 & Scd & 0.15 & 9.95 & 7.77\\
231.67357 & 40.55709 & F15248+4044 & 0.4 & 0.008743 & & 9.70 & 9.43 & 9.41 & & 10.16 & Scd & 0.7 & 10.60 & 8.31\\
226.95305 & 54.74415 & F15064+5456 & 0.7 & 0.01043 & & 10.00 & & & & 10.16 & Scd & 0.4 & 10.60 & 7.81\\
126.57545 & 22.88144 & F08233+2303 & 1.3 & 0.01794 & & 10.20 & 10.13 & & & 9.81 & Scd & 0.3 & 10.25& 8.99\\
231.43570 & 52.44624 & F15243+5237 & 0.3 & 0.01948 & 10.19 & & 9.81 & & & 10.44 & Sab & 0.2 & 11.01 & 8.64\\
237.68201 & 55.60954 & F15495+5545 & 0.2 & 0.03974 & 10.55 & & 10.52 & 10.20 & & 10.70 & Sab & 0.0 & 11.27 & 9.34\\
66.76954 & -49.12881  & F04257-4913 & 0.8 & 0.05828 & & & 10.62 & 11.55 & & 10.85 & Sab & 0.0 & 11.42 & 9.43\\
\\ \hline \\~\\
\end{tabular}
Positions given are from the ERCSC. Where the sources have a well known name this is given beneath the {\it IRAS\/} name. All luminosities are measured in solar luminosities.
\end{tiny}
\end{table*}

\begin{figure}
\centering
\includegraphics[angle=0,scale=0.4]{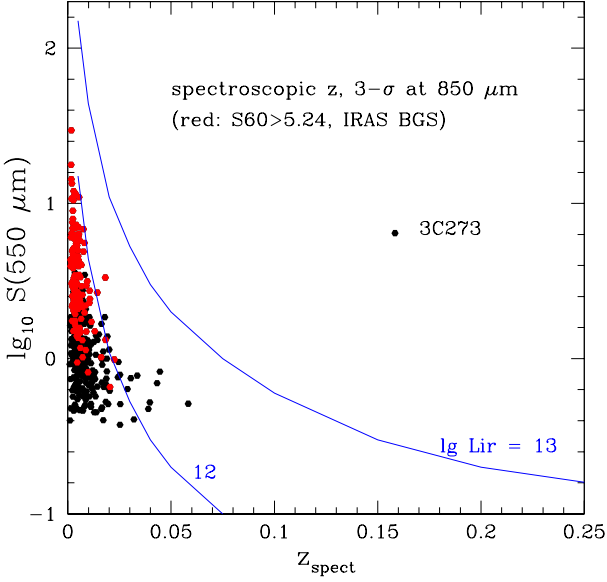}
\caption{ \getsymbol{HFI:center:frequency:545GHz} GHz flux density versus redshift, showing loci for Arp220 template with $L_{\rm IR}$ =10$^{12}$ and 10$^{13} {\rm L}_{\odot}$. Red dots are those galaxies in the IRAS Bright Galaxy Survey (BGS). We show here that the BGS does not sample as wide a range of galaxy properties as the Planck-IIFSCz sample discussed here.}
\end{figure}

\begin{figure}
\centering
\includegraphics[angle=0,scale=0.4]{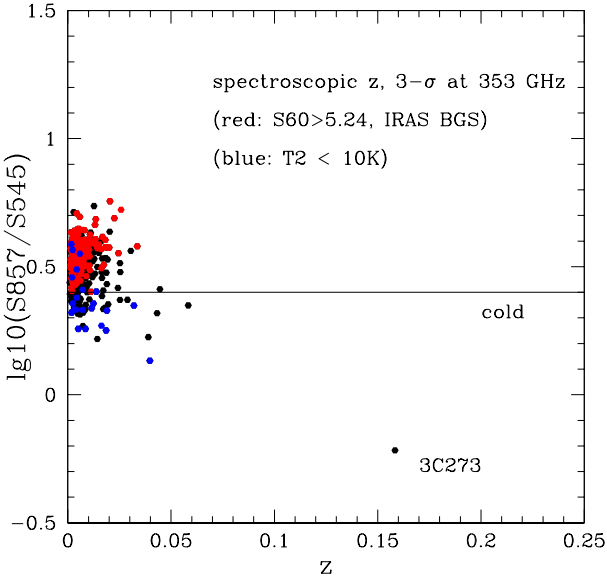}
\caption{ S$_{\getsymbol{HFI:center:frequency:857GHz}}$/ S$_{\getsymbol{HFI:center:frequency:545GHz}}$ colour-ratio versus redshift. Points in red are {\it IRAS\/}
BGS sources, points in blue are those sources identified as having $T_2<10\,$K on the basis of parametric fits.}
\end{figure}

\begin{figure}
\centering
\includegraphics[angle=0,scale=0.4]{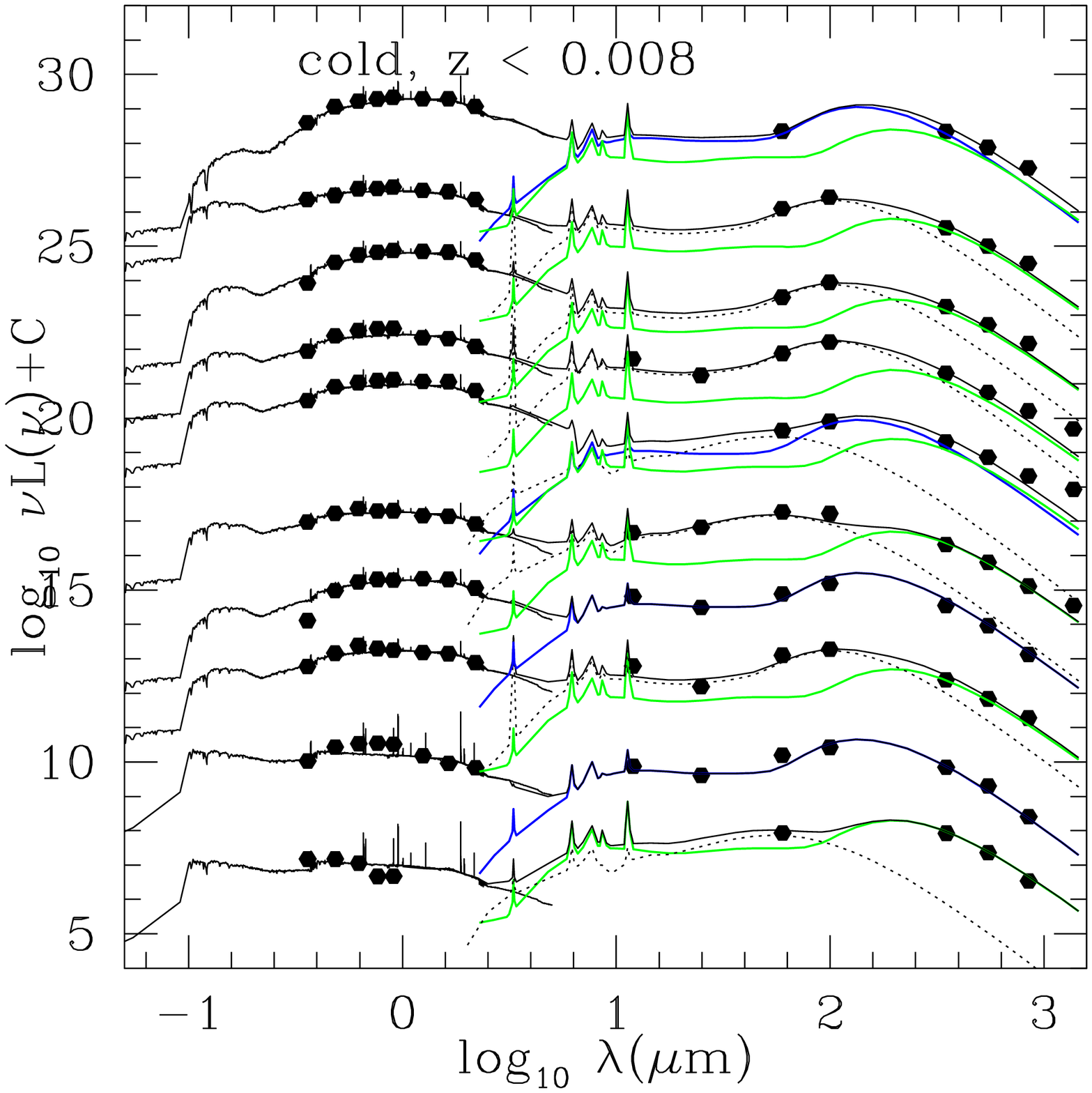}
\includegraphics[angle=0,scale=0.4]{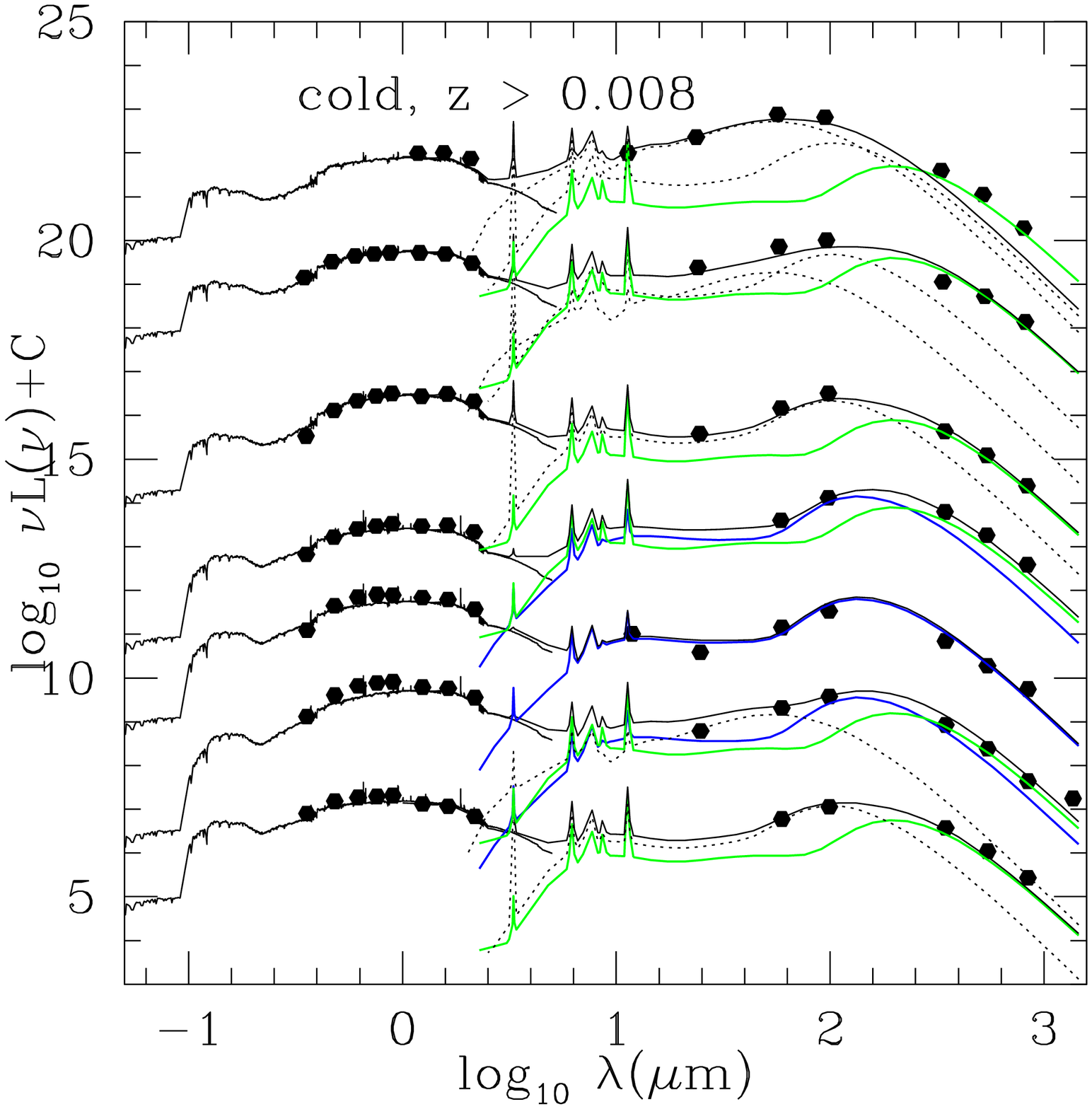}
\caption{Template fits for cool ERCSC galaxies (log$_{10}$(S$_{ \getsymbol{HFI:center:frequency:857GHz}}$/S$_{\getsymbol{HFI:center:frequency:545GHz}}$) $<$ 0.4).
Blue curve: cirrus with $\phi$ = 1, green curve: cirrus with $\phi$ = 0.1.}
\end{figure}

\section{Conclusions}\label{sec:Conclusions}

Our studies of nearby galaxies detected by \Planck\ have shown evidence for colder dust than has previously been found in local galaxies. The temperature of this dust, however, is similar to that found in the solar neighbourhood. We also find that some local galaxies are both luminous and cool, with properties similar to those of the distant SMGs uncovered in deep submm surveys. This suggests that previous studies of dust in local galaxies have been biased away from such luminous cool objects. We also find that the dust SEDs in most galaxies are better described by parametric models containing two dust components, one warm and one cold, with the cold component reaching temperatures as low as $10\,$K. Some objects have SEDs dominated by this cold material. These conclusions are based on both parametric fits and by detailed fitting of radiative transfer derived physical templates to the SEDs. However, other physical or parametric descriptions of dust, for example where $\beta$ varies with wavelength, might lead to different results regarding this very cold component. This paper represents the first exploitation of \Planck\ data for the study of a large sample of galaxies in the local Universe. As such it indicates both the benefits and hazards of the ERCSC for this work, but it also clearly demonstrates the tremendous potential of \Planck\ data for the study of dust in galaxies.

\begin{acknowledgements}
This research has made use of the NASA/IPAC Extragalactic Database (NED) which is
operated by the Jet Propulsion Laboratory, California Institute of Technology,
under contract with the National Aeronautics and Space Administration.  Use was
also made of data from the Sloan Digital Sky Survey (SDSS) and the Two Micron
All Sky Survey (2MASS).
Funding for the SDSS and SDSS-II has been provided by the Alfred P. Sloan
Foundation, the Participating Institutions, the National Science Foundation,
the U.S. Department of Energy, the National Aeronautics and Space Administration,
the Japanese Monbukagakusho, the Max Planck Society, and the Higher Education
Funding Council for England. The SDSS Web Site is {\tt http://www.sdss.org/}\,.
2MASS is a joint project of the University of Massachusetts and the Infrared
Processing and Analysis Center/California Institute of Technology, funded by the
National Aeronautics and Space Administration and the National Science Foundation.
A description of the Planck Collaboration and a list of its members, indicating which technical or scientific activities they have been involved in, can be found at \tt{http://www.rssd.esa.int/index.php?project=PLANCK\&page=
Planck\_Collaboration}
\end{acknowledgements}

\bibliographystyle{aa}
\bibliography{Planck_bib,dlcrefs}

\end{document}

%% file: Planck.tex
\def\setsymbol#1#2{\expandafter\def\csname #1\endcsname{#2}}
\def\getsymbol#1{\csname #1\endcsname}

\def\Planck{{\it Planck\/}}

\newbox\tablebox    \newdimen\tablewidth
\def\leaderfil{\leaders\hbox to 5pt{\hss.\hss}\hfil}
\def\tablenote#1 #2\par{\begingroup \parindent=0.8em
    \abovedisplayshortskip=0pt\belowdisplayshortskip=0pt
    \noindent
    $$\hss\vbox{\hsize\tablewidth \hangindent=\parindent \hangafter=1 \noindent
    \hbox to \parindent{\sup{\rm #1}\hss}\strut#2\strut\par}\hss$$
    \endgroup}

\def\L2{\ifmmode L_2\else $L_2$\fi}

\def\DeltaT{\ifmmode \Delta T\else $\Delta T$\fi}
\def\deltat{\ifmmode \Delta t\else $\Delta t$\fi}
\def\fknee{\ifmmode f_{\rm knee}\else $f_{\rm knee}$\fi}
\def\Fmax{\ifmmode F_{\rm max}\else $F_{\rm max}$\fi}
\def\solar{\ifmmode{\rm M}_{\mathord\odot}\else${\rm M}_{\mathord\odot}$\fi}

\def\inv{\ifmmode^{-1}\else$^{-1}$\fi}
\def\mo{\ifmmode^{-1}\else$^{-1}$\fi}
\def\sup#1{\ifmmode ^{\rm #1}\else $^{\rm #1}$\fi}
\def\expo#1{\ifmmode \times 10^{#1}\else $\times 10^{#1}$\fi}
\def\,{\thinspace}
\def\lsim{\mathrel{\raise .4ex\hbox{\rlap{$<$}\lower 1.2ex\hbox{$\sim$}}}}
\def\gsim{\mathrel{\raise .4ex\hbox{\rlap{$>$}\lower 1.2ex\hbox{$\sim$}}}}

\def\simprop{\mathrel{\raise .4ex\hbox{\rlap{$\propto$}\lower 1.2ex\hbox{$\sim$}}}}
\def\deg{\ifmmode^\circ\else$^\circ$\fi}
\def\pdeg{\ifmmode $\setbox0=\hbox{$^{\circ}$}\rlap{\hskip.11\wd0 .}$^{\circ}
          \else \setbox0=\hbox{$^{\circ}$}\rlap{\hskip.11\wd0 .}$^{\circ}$\fi}
\def\arcs{\ifmmode {^{\scriptstyle\prime\prime}}
          \else $^{\scriptstyle\prime\prime}$\fi}
\def\arcm{\ifmmode {^{\scriptstyle\prime}}
          \else $^{\scriptstyle\prime}$\fi}
\newdimen\sa  \newdimen\sb
\def\parcs{\sa=.07em \sb=.03em
     \ifmmode \hbox{\rlap{.}}^{\scriptstyle\prime\kern -\sb\prime}\hbox{\kern -\sa}
     \else \rlap{.}$^{\scriptstyle\prime\kern -\sb\prime}$\kern -\sa\fi}
\def\parcm{\sa=.08em \sb=.03em
     \ifmmode \hbox{\rlap{.}\kern\sa}^{\scriptstyle\prime}\hbox{\kern-\sb}
     \else \rlap{.}\kern\sa$^{\scriptstyle\prime}$\kern-\sb\fi}
\def\ra[#1 #2 #3.#4]{#1\sup{h}#2\sup{m}#3\sup{s}\llap.#4}
\def\dec[#1 #2 #3.#4]{#1\deg#2\arcm#3\arcs\llap.#4}
\def\deco[#1 #2 #3]{#1\deg#2\arcm#3\arcs}
\def\rra[#1 #2]{#1\sup{h}#2\sup{m}}

\def\dots{\relax\ifmmode \ldots\else $\ldots$\fi}
\def\WHzsr{\ifmmode $W\,Hz\mo\,sr\mo$\else W\,Hz\mo\,sr\mo\fi}
\def\mHz{\ifmmode $\,mHz$\else \,mHz\fi}
\def\GHz{\ifmmode $\,GHz$\else \,GHz\fi}
\def\mKs{\ifmmode $\,mK\,s$^{1/2}\else \,mK\,s$^{1/2}$\fi}
\def\muKs{\ifmmode \,\mu$K\,s$^{1/2}\else \,$\mu$K\,s$^{1/2}$\fi}
\def\muKRJs{\ifmmode \,\mu$K$_{\rm RJ}$\,s$^{1/2}\else \,$\mu$K$_{\rm RJ}$\,s$^{1/2}$\fi}
\def\muKHz{\ifmmode \,\mu$K\,Hz$^{-1/2}\else \,$\mu$K\,Hz$^{-1/2}$\fi}
\def\MJysr{\ifmmode \,$MJy\,sr\mo$\else \,MJy\,sr\mo\fi}
\def\MJysrmK{\ifmmode \,$MJy\,sr\mo$\,mK$_{\rm CMB}\mo\else \,MJy\,sr\mo\,mK$_{\rm CMB}\mo$\fi}
\def\microns{\ifmmode \,\mu$m$\else \,$\mu$m\fi}

\def\muK{\ifmmode \,\mu$K$\else \,$\mu$\hbox{K}\fi}
\def\microK{\ifmmode \,\mu$K$\else \,$\mu$\hbox{K}\fi}
\def\muW{\ifmmode \,\mu$W$\else \,$\mu$\hbox{W}\fi}
\def\kms{\ifmmode $\,km\,s$^{-1}\else \,km\,s$^{-1}$\fi}
\def\kmsMpc{\ifmmode $\,\kms\,Mpc\mo$\else \,\kms\,Mpc\mo\fi}
\setsymbol{LFI:center:frequency:70GHz:units}{70.3\,GHz}
\setsymbol{LFI:center:frequency:44GHz:units}{44.1\,GHz}
\setsymbol{LFI:center:frequency:30GHz:units}{28.5\,GHz}

\setsymbol{LFI:center:frequency:70GHz}{70.3}
\setsymbol{LFI:center:frequency:44GHz}{44.1}
\setsymbol{LFI:center:frequency:30GHz}{28.5}

\setsymbol{LFI:center:frequency:LFI18:Rad:M:units}{71.7\GHz}
\setsymbol{LFI:center:frequency:LFI19:Rad:M:units}{67.5\GHz}
\setsymbol{LFI:center:frequency:LFI20:Rad:M:units}{69.2\GHz}
\setsymbol{LFI:center:frequency:LFI21:Rad:M:units}{70.4\GHz}
\setsymbol{LFI:center:frequency:LFI22:Rad:M:units}{71.5\GHz}
\setsymbol{LFI:center:frequency:LFI23:Rad:M:units}{70.8\GHz}
\setsymbol{LFI:center:frequency:LFI24:Rad:M:units}{44.4\GHz}
\setsymbol{LFI:center:frequency:LFI25:Rad:M:units}{44.0\GHz}
\setsymbol{LFI:center:frequency:LFI26:Rad:M:units}{43.9\GHz}
\setsymbol{LFI:center:frequency:LFI27:Rad:M:units}{28.3\GHz}
\setsymbol{LFI:center:frequency:LFI28:Rad:M:units}{28.8\GHz}
\setsymbol{LFI:center:frequency:LFI18:Rad:S:units}{70.1\GHz}
\setsymbol{LFI:center:frequency:LFI19:Rad:S:units}{69.6\GHz}
\setsymbol{LFI:center:frequency:LFI20:Rad:S:units}{69.5\GHz}
\setsymbol{LFI:center:frequency:LFI21:Rad:S:units}{69.5\GHz}
\setsymbol{LFI:center:frequency:LFI22:Rad:S:units}{72.8\GHz}
\setsymbol{LFI:center:frequency:LFI23:Rad:S:units}{71.3\GHz}
\setsymbol{LFI:center:frequency:LFI24:Rad:S:units}{44.1\GHz}
\setsymbol{LFI:center:frequency:LFI25:Rad:S:units}{44.1\GHz}
\setsymbol{LFI:center:frequency:LFI26:Rad:S:units}{44.1\GHz}
\setsymbol{LFI:center:frequency:LFI27:Rad:S:units}{28.5\GHz}
\setsymbol{LFI:center:frequency:LFI28:Rad:S:units}{28.2\GHz}

\setsymbol{LFI:center:frequency:LFI18:Rad:M}{71.7}
\setsymbol{LFI:center:frequency:LFI19:Rad:M}{67.5}
\setsymbol{LFI:center:frequency:LFI20:Rad:M}{69.2}
\setsymbol{LFI:center:frequency:LFI21:Rad:M}{70.4}
\setsymbol{LFI:center:frequency:LFI22:Rad:M}{71.5}
\setsymbol{LFI:center:frequency:LFI23:Rad:M}{70.8}
\setsymbol{LFI:center:frequency:LFI24:Rad:M}{44.4}
\setsymbol{LFI:center:frequency:LFI25:Rad:M}{44.0}
\setsymbol{LFI:center:frequency:LFI26:Rad:M}{43.9}
\setsymbol{LFI:center:frequency:LFI27:Rad:M}{28.3}
\setsymbol{LFI:center:frequency:LFI28:Rad:M}{28.8}
\setsymbol{LFI:center:frequency:LFI18:Rad:S}{70.1}
\setsymbol{LFI:center:frequency:LFI19:Rad:S}{69.6}
\setsymbol{LFI:center:frequency:LFI20:Rad:S}{69.5}
\setsymbol{LFI:center:frequency:LFI21:Rad:S}{69.5}
\setsymbol{LFI:center:frequency:LFI22:Rad:S}{72.8}
\setsymbol{LFI:center:frequency:LFI23:Rad:S}{71.3}
\setsymbol{LFI:center:frequency:LFI24:Rad:S}{44.1}
\setsymbol{LFI:center:frequency:LFI25:Rad:S}{44.1}
\setsymbol{LFI:center:frequency:LFI26:Rad:S}{44.1}
\setsymbol{LFI:center:frequency:LFI27:Rad:S}{28.5}
\setsymbol{LFI:center:frequency:LFI28:Rad:S}{28.2}

\setsymbol{LFI:white:noise:sensitivity:70GHz:units}{152.6\muKs}
\setsymbol{LFI:white:noise:sensitivity:44GHz:units}{173.1\muKs}
\setsymbol{LFI:white:noise:sensitivity:30GHz:units}{146.8\muKs}

\setsymbol{LFI:white:noise:sensitivity:70GHz}{152.6}
\setsymbol{LFI:white:noise:sensitivity:44GHz}{173.1}
\setsymbol{LFI:white:noise:sensitivity:30GHz}{146.8}

\setsymbol{LFI:white:noise:sensitivity:LFI18:Rad:M:units}{512.0\muKs}
\setsymbol{LFI:white:noise:sensitivity:LFI19:Rad:M:units}{581.4\muKs}
\setsymbol{LFI:white:noise:sensitivity:LFI20:Rad:M:units}{590.8\muKs}
\setsymbol{LFI:white:noise:sensitivity:LFI21:Rad:M:units}{455.2\muKs}
\setsymbol{LFI:white:noise:sensitivity:LFI22:Rad:M:units}{492.0\muKs}
\setsymbol{LFI:white:noise:sensitivity:LFI23:Rad:M:units}{507.7\muKs}
\setsymbol{LFI:white:noise:sensitivity:LFI24:Rad:M:units}{462.2\muKs}
\setsymbol{LFI:white:noise:sensitivity:LFI25:Rad:M:units}{413.6\muKs}
\setsymbol{LFI:white:noise:sensitivity:LFI26:Rad:M:units}{478.6\muKs}
\setsymbol{LFI:white:noise:sensitivity:LFI27:Rad:M:units}{277.7\muKs}
\setsymbol{LFI:white:noise:sensitivity:LFI28:Rad:M:units}{312.3\muKs}
\setsymbol{LFI:white:noise:sensitivity:LFI18:Rad:S:units}{465.7\muKs}
\setsymbol{LFI:white:noise:sensitivity:LFI19:Rad:S:units}{555.6\muKs}
\setsymbol{LFI:white:noise:sensitivity:LFI20:Rad:S:units}{623.2\muKs}
\setsymbol{LFI:white:noise:sensitivity:LFI21:Rad:S:units}{564.1\muKs}
\setsymbol{LFI:white:noise:sensitivity:LFI22:Rad:S:units}{534.4\muKs}
\setsymbol{LFI:white:noise:sensitivity:LFI23:Rad:S:units}{542.4\muKs}
\setsymbol{LFI:white:noise:sensitivity:LFI24:Rad:S:units}{399.2\muKs}
\setsymbol{LFI:white:noise:sensitivity:LFI25:Rad:S:units}{392.6\muKs}
\setsymbol{LFI:white:noise:sensitivity:LFI26:Rad:S:units}{418.6\muKs}
\setsymbol{LFI:white:noise:sensitivity:LFI27:Rad:S:units}{302.9\muKs}
\setsymbol{LFI:white:noise:sensitivity:LFI28:Rad:S:units}{285.3\muKs}

\setsymbol{LFI:white:noise:sensitivity:LFI18:Rad:M}{512.0}
\setsymbol{LFI:white:noise:sensitivity:LFI19:Rad:M}{581.4}
\setsymbol{LFI:white:noise:sensitivity:LFI20:Rad:M}{590.8}
\setsymbol{LFI:white:noise:sensitivity:LFI21:Rad:M}{455.2}
\setsymbol{LFI:white:noise:sensitivity:LFI22:Rad:M}{492.0}
\setsymbol{LFI:white:noise:sensitivity:LFI23:Rad:M}{507.7}
\setsymbol{LFI:white:noise:sensitivity:LFI24:Rad:M}{462.2}
\setsymbol{LFI:white:noise:sensitivity:LFI25:Rad:M}{413.6}
\setsymbol{LFI:white:noise:sensitivity:LFI26:Rad:M}{478.6}
\setsymbol{LFI:white:noise:sensitivity:LFI27:Rad:M}{277.7}
\setsymbol{LFI:white:noise:sensitivity:LFI28:Rad:M}{312.3}
\setsymbol{LFI:white:noise:sensitivity:LFI18:Rad:S}{465.7}
\setsymbol{LFI:white:noise:sensitivity:LFI19:Rad:S}{555.6}
\setsymbol{LFI:white:noise:sensitivity:LFI20:Rad:S}{623.2}
\setsymbol{LFI:white:noise:sensitivity:LFI21:Rad:S}{564.1}
\setsymbol{LFI:white:noise:sensitivity:LFI22:Rad:S}{534.4}
\setsymbol{LFI:white:noise:sensitivity:LFI23:Rad:S}{542.4}
\setsymbol{LFI:white:noise:sensitivity:LFI24:Rad:S}{399.2}
\setsymbol{LFI:white:noise:sensitivity:LFI25:Rad:S}{392.6}
\setsymbol{LFI:white:noise:sensitivity:LFI26:Rad:S}{418.6}
\setsymbol{LFI:white:noise:sensitivity:LFI27:Rad:S}{302.9}
\setsymbol{LFI:white:noise:sensitivity:LFI28:Rad:S}{285.3}

\setsymbol{LFI:knee:frequency:70GHz:units}{29.5\mHz}
\setsymbol{LFI:knee:frequency:44GHz:units}{56.2\mHz}
\setsymbol{LFI:knee:frequency:30GHz:units}{113.7\mHz}

\setsymbol{LFI:knee:frequency:70GHz}{29.5}
\setsymbol{LFI:knee:frequency:44GHz}{56.2}
\setsymbol{LFI:knee:frequency:30GHz}{113.7}

\setsymbol{LFI:knee:frequency:LFI18:Rad:M:units}{16.3\mHz}
\setsymbol{LFI:knee:frequency:LFI19:Rad:M:units}{15.1\mHz}
\setsymbol{LFI:knee:frequency:LFI20:Rad:M:units}{18.7\mHz}
\setsymbol{LFI:knee:frequency:LFI21:Rad:M:units}{37.2\mHz}
\setsymbol{LFI:knee:frequency:LFI22:Rad:M:units}{12.7\mHz}
\setsymbol{LFI:knee:frequency:LFI23:Rad:M:units}{34.6\mHz}
\setsymbol{LFI:knee:frequency:LFI24:Rad:M:units}{46.2\mHz}
\setsymbol{LFI:knee:frequency:LFI25:Rad:M:units}{24.9\mHz}
\setsymbol{LFI:knee:frequency:LFI26:Rad:M:units}{67.6\mHz}
\setsymbol{LFI:knee:frequency:LFI27:Rad:M:units}{187.4\mHz}
\setsymbol{LFI:knee:frequency:LFI28:Rad:M:units}{122.2\mHz}
\setsymbol{LFI:knee:frequency:LFI18:Rad:S:units}{17.7\mHz}
\setsymbol{LFI:knee:frequency:LFI19:Rad:S:units}{22.0\mHz}
\setsymbol{LFI:knee:frequency:LFI20:Rad:S:units}{8.7\mHz}
\setsymbol{LFI:knee:frequency:LFI21:Rad:S:units}{25.9\mHz}
\setsymbol{LFI:knee:frequency:LFI22:Rad:S:units}{15.8\mHz}
\setsymbol{LFI:knee:frequency:LFI23:Rad:S:units}{129.8\mHz}
\setsymbol{LFI:knee:frequency:LFI24:Rad:S:units}{100.9\mHz}
\setsymbol{LFI:knee:frequency:LFI25:Rad:S:units}{38.9\mHz}
\setsymbol{LFI:knee:frequency:LFI26:Rad:S:units}{58.9\mHz}
\setsymbol{LFI:knee:frequency:LFI27:Rad:S:units}{104.4\mHz}
\setsymbol{LFI:knee:frequency:LFI28:Rad:S:units}{40.7\mHz}

\setsymbol{LFI:knee:frequency:LFI18:Rad:M}{16.3}
\setsymbol{LFI:knee:frequency:LFI19:Rad:M}{15.1}
\setsymbol{LFI:knee:frequency:LFI20:Rad:M}{18.7}
\setsymbol{LFI:knee:frequency:LFI21:Rad:M}{37.2}
\setsymbol{LFI:knee:frequency:LFI22:Rad:M}{12.7}
\setsymbol{LFI:knee:frequency:LFI23:Rad:M}{34.6}
\setsymbol{LFI:knee:frequency:LFI24:Rad:M}{46.2}
\setsymbol{LFI:knee:frequency:LFI25:Rad:M}{24.9}
\setsymbol{LFI:knee:frequency:LFI26:Rad:M}{67.6}
\setsymbol{LFI:knee:frequency:LFI27:Rad:M}{187.4}
\setsymbol{LFI:knee:frequency:LFI28:Rad:M}{122.2}
\setsymbol{LFI:knee:frequency:LFI18:Rad:S}{17.7}
\setsymbol{LFI:knee:frequency:LFI19:Rad:S}{22.0}
\setsymbol{LFI:knee:frequency:LFI20:Rad:S}{8.7}
\setsymbol{LFI:knee:frequency:LFI21:Rad:S}{25.9}
\setsymbol{LFI:knee:frequency:LFI22:Rad:S}{15.8}
\setsymbol{LFI:knee:frequency:LFI23:Rad:S}{129.8}
\setsymbol{LFI:knee:frequency:LFI24:Rad:S}{100.9}
\setsymbol{LFI:knee:frequency:LFI25:Rad:S}{38.9}
\setsymbol{LFI:knee:frequency:LFI26:Rad:S}{58.9}
\setsymbol{LFI:knee:frequency:LFI27:Rad:S}{104.4}
\setsymbol{LFI:knee:frequency:LFI28:Rad:S}{40.7}

\setsymbol{LFI:slope:70GHz:units}{$-1.03$\mHz}
\setsymbol{LFI:slope:44GHz:units}{$-0.89$\mHz}
\setsymbol{LFI:slope:30GHz:units}{$-0.87$\mHz}

\setsymbol{LFI:slope:70GHz}{$-1.03$}
\setsymbol{LFI:slope:44GHz}{$-0.89$}
\setsymbol{LFI:slope:30GHz}{$-0.87$}

\setsymbol{LFI:slope:LFI18:Rad:M:units}{$-1.04$\mHz}
\setsymbol{LFI:slope:LFI19:Rad:M:units}{$-1.09$\mHz}
\setsymbol{LFI:slope:LFI20:Rad:M:units}{$-0.69$\mHz}
\setsymbol{LFI:slope:LFI21:Rad:M:units}{$-1.56$\mHz}
\setsymbol{LFI:slope:LFI22:Rad:M:units}{$-1.01$\mHz}
\setsymbol{LFI:slope:LFI23:Rad:M:units}{$-0.96$\mHz}
\setsymbol{LFI:slope:LFI24:Rad:M:units}{$-0.83$\mHz}
\setsymbol{LFI:slope:LFI25:Rad:M:units}{$-0.91$\mHz}
\setsymbol{LFI:slope:LFI26:Rad:M:units}{$-0.95$\mHz}
\setsymbol{LFI:slope:LFI27:Rad:M:units}{$-0.87$\mHz}
\setsymbol{LFI:slope:LFI28:Rad:M:units}{$-0.88$\mHz}
\setsymbol{LFI:slope:LFI18:Rad:S:units}{$-1.15$\mHz}
\setsymbol{LFI:slope:LFI19:Rad:S:units}{$-1.00$\mHz}
\setsymbol{LFI:slope:LFI20:Rad:S:units}{$-0.95$\mHz}
\setsymbol{LFI:slope:LFI21:Rad:S:units}{$-0.92$\mHz}
\setsymbol{LFI:slope:LFI22:Rad:S:units}{$-1.01$\mHz}
\setsymbol{LFI:slope:LFI23:Rad:S:units}{$-0.95$\mHz}
\setsymbol{LFI:slope:LFI24:Rad:S:units}{$-0.73$\mHz}
\setsymbol{LFI:slope:LFI25:Rad:S:units}{$-1.16$\mHz}
\setsymbol{LFI:slope:LFI26:Rad:S:units}{$-0.79$\mHz}
\setsymbol{LFI:slope:LFI27:Rad:S:units}{$-0.82$\mHz}
\setsymbol{LFI:slope:LFI28:Rad:S:units}{$-0.91$\mHz}

\setsymbol{LFI:slope:LFI18:Rad:M}{$-1.04$}
\setsymbol{LFI:slope:LFI19:Rad:M}{$-1.09$}
\setsymbol{LFI:slope:LFI20:Rad:M}{$-0.69$}
\setsymbol{LFI:slope:LFI21:Rad:M}{$-1.56$}
\setsymbol{LFI:slope:LFI22:Rad:M}{$-1.01$}
\setsymbol{LFI:slope:LFI23:Rad:M}{$-0.96$}
\setsymbol{LFI:slope:LFI24:Rad:M}{$-0.83$}
\setsymbol{LFI:slope:LFI25:Rad:M}{$-0.91$}
\setsymbol{LFI:slope:LFI26:Rad:M}{$-0.95$}
\setsymbol{LFI:slope:LFI27:Rad:M}{$-0.87$}
\setsymbol{LFI:slope:LFI28:Rad:M}{$-0.88$}
\setsymbol{LFI:slope:LFI18:Rad:S}{$-1.15$}
\setsymbol{LFI:slope:LFI19:Rad:S}{$-1.00$}
\setsymbol{LFI:slope:LFI20:Rad:S}{$-0.95$}
\setsymbol{LFI:slope:LFI21:Rad:S}{$-0.92$}
\setsymbol{LFI:slope:LFI22:Rad:S}{$-1.01$}
\setsymbol{LFI:slope:LFI23:Rad:S}{$-0.95$}
\setsymbol{LFI:slope:LFI24:Rad:S}{$-0.73$}
\setsymbol{LFI:slope:LFI25:Rad:S}{$-1.16$}
\setsymbol{LFI:slope:LFI26:Rad:S}{$-0.79$}
\setsymbol{LFI:slope:LFI27:Rad:S}{$-0.82$}
\setsymbol{LFI:slope:LFI28:Rad:S}{$-0.91$}

\setsymbol{LFI:FWHM:70GHz:units}{13\parcm01}
\setsymbol{LFI:FWHM:44GHz:units}{27\parcm92}
\setsymbol{LFI:FWHM:30GHz:units}{32\parcm65}

\setsymbol{LFI:FWHM:70GHz}{13.01}
\setsymbol{LFI:FWHM:44GHz}{27.92}
\setsymbol{LFI:FWHM:30GHz}{32.65}

\setsymbol{LFI:FWHM:LFI18:units}{13\parcm39}
\setsymbol{LFI:FWHM:LFI19:units}{13\parcm01}
\setsymbol{LFI:FWHM:LFI20:units}{12\parcm75}
\setsymbol{LFI:FWHM:LFI21:units}{12\parcm74}
\setsymbol{LFI:FWHM:LFI22:units}{12\parcm87}
\setsymbol{LFI:FWHM:LFI23:units}{13\parcm27}
\setsymbol{LFI:FWHM:LFI24:units}{22\parcm98}
\setsymbol{LFI:FWHM:LFI25:units}{30\parcm46}
\setsymbol{LFI:FWHM:LFI26:units}{30\parcm31}
\setsymbol{LFI:FWHM:LFI27:units}{32\parcm65}
\setsymbol{LFI:FWHM:LFI28:units}{32\parcm66}

\setsymbol{LFI:FWHM:LFI18}{13.39}
\setsymbol{LFI:FWHM:LFI19}{13.01}
\setsymbol{LFI:FWHM:LFI20}{12.75}
\setsymbol{LFI:FWHM:LFI21}{12.74}
\setsymbol{LFI:FWHM:LFI22}{12.87}
\setsymbol{LFI:FWHM:LFI23}{13.27}
\setsymbol{LFI:FWHM:LFI24}{22.98}
\setsymbol{LFI:FWHM:LFI25}{30.46}
\setsymbol{LFI:FWHM:LFI26}{30.31}
\setsymbol{LFI:FWHM:LFI27}{32.65}
\setsymbol{LFI:FWHM:LFI28}{32.66}

\setsymbol{LFI:FWHM:uncertainty:LFI18:units}{0.170\arcm}
\setsymbol{LFI:FWHM:uncertainty:LFI19:units}{0.174\arcm}
\setsymbol{LFI:FWHM:uncertainty:LFI20:units}{0.170\arcm}
\setsymbol{LFI:FWHM:uncertainty:LFI21:units}{0.156\arcm}
\setsymbol{LFI:FWHM:uncertainty:LFI22:units}{0.164\arcm}
\setsymbol{LFI:FWHM:uncertainty:LFI23:units}{0.171\arcm}
\setsymbol{LFI:FWHM:uncertainty:LFI24:units}{0.652\arcm}
\setsymbol{LFI:FWHM:uncertainty:LFI25:units}{1.075\arcm}
\setsymbol{LFI:FWHM:uncertainty:LFI26:units}{1.131\arcm}
\setsymbol{LFI:FWHM:uncertainty:LFI27:units}{1.266\arcm}
\setsymbol{LFI:FWHM:uncertainty:LFI28:units}{1.287\arcm}

\setsymbol{LFI:FWHM:uncertainty:LFI18}{0.170}
\setsymbol{LFI:FWHM:uncertainty:LFI19}{0.174}
\setsymbol{LFI:FWHM:uncertainty:LFI20}{0.170}
\setsymbol{LFI:FWHM:uncertainty:LFI21}{0.156}
\setsymbol{LFI:FWHM:uncertainty:LFI22}{0.164}
\setsymbol{LFI:FWHM:uncertainty:LFI23}{0.171}
\setsymbol{LFI:FWHM:uncertainty:LFI24}{0.652}
\setsymbol{LFI:FWHM:uncertainty:LFI25}{1.075}
\setsymbol{LFI:FWHM:uncertainty:LFI26}{1.131}
\setsymbol{LFI:FWHM:uncertainty:LFI27}{1.266}
\setsymbol{LFI:FWHM:uncertainty:LFI28}{1.287}

\setsymbol{HFI:center:frequency:100GHz:units}{100\,GHz}
\setsymbol{HFI:center:frequency:143GHz:units}{143\,GHz}
\setsymbol{HFI:center:frequency:217GHz:units}{217\,GHz}
\setsymbol{HFI:center:frequency:353GHz:units}{353\,GHz}
\setsymbol{HFI:center:frequency:545GHz:units}{545\,GHz}
\setsymbol{HFI:center:frequency:857GHz:units}{857\,GHz}

\setsymbol{HFI:center:frequency:100GHz}{100}
\setsymbol{HFI:center:frequency:143GHz}{143}
\setsymbol{HFI:center:frequency:217GHz}{217}
\setsymbol{HFI:center:frequency:353GHz}{353}
\setsymbol{HFI:center:frequency:545GHz}{545}
\setsymbol{HFI:center:frequency:857GHz}{857}

\setsymbol{HFI:Ndetectors:100GHz}{8}
\setsymbol{HFI:Ndetectors:143GHz}{11}
\setsymbol{HFI:Ndetectors:217GHz}{12}
\setsymbol{HFI:Ndetectors:353GHz}{12}
\setsymbol{HFI:Ndetectors:545GHz}{3}
\setsymbol{HFI:Ndetectors:857GHz}{4}

\setsymbol{HFI:FWHM:Maps:100GHz:units}{9\parcm88}
\setsymbol{HFI:FWHM:Maps:143GHz:units}{7\parcm18}
\setsymbol{HFI:FWHM:Maps:217GHz:units}{4\parcm87}
\setsymbol{HFI:FWHM:Maps:353GHz:units}{4\parcm65}
\setsymbol{HFI:FWHM:Maps:545GHz:units}{4\parcm72}
\setsymbol{HFI:FWHM:Maps:857GHz:units}{4\parcm39}
\setsymbol{HFI:FWHM:Maps:100GHz}{9.88}
\setsymbol{HFI:FWHM:Maps:143GHz}{7.18}
\setsymbol{HFI:FWHM:Maps:217GHz}{4.87}
\setsymbol{HFI:FWHM:Maps:353GHz}{4.65}
\setsymbol{HFI:FWHM:Maps:545GHz}{4.72}
\setsymbol{HFI:FWHM:Maps:857GHz}{4.39}

\setsymbol{HFI:beam:ellipticity:Maps:100GHz}{1.15}
\setsymbol{HFI:beam:ellipticity:Maps:143GHz}{1.01}
\setsymbol{HFI:beam:ellipticity:Maps:217GHz}{1.06}
\setsymbol{HFI:beam:ellipticity:Maps:353GHz}{1.05}
\setsymbol{HFI:beam:ellipticity:Maps:545GHz}{1.14}
\setsymbol{HFI:beam:ellipticity:Maps:857GHz}{1.19}

\setsymbol{HFI:FWHM:Mars:100GHz:units}{9\parcm37}
\setsymbol{HFI:FWHM:Mars:143GHz:units}{7\parcm04}
\setsymbol{HFI:FWHM:Mars:217GHz:units}{4\parcm68}
\setsymbol{HFI:FWHM:Mars:353GHz:units}{4\parcm43}
\setsymbol{HFI:FWHM:Mars:545GHz:units}{3\parcm80}
\setsymbol{HFI:FWHM:Mars:857GHz:units}{3\parcm67}

\setsymbol{HFI:FWHM:Mars:100GHz}{9.37}
\setsymbol{HFI:FWHM:Mars:143GHz}{7.04}
\setsymbol{HFI:FWHM:Mars:217GHz}{4.68}
\setsymbol{HFI:FWHM:Mars:353GHz}{4.43}
\setsymbol{HFI:FWHM:Mars:545GHz}{3.80}
\setsymbol{HFI:FWHM:Mars:857GHz}{3.67}

\setsymbol{HFI:beam:ellipticity:Mars:100GHz}{1.18}
\setsymbol{HFI:beam:ellipticity:Mars:143GHz}{1.03}
\setsymbol{HFI:beam:ellipticity:Mars:217GHz}{1.14}
\setsymbol{HFI:beam:ellipticity:Mars:353GHz}{1.09}
\setsymbol{HFI:beam:ellipticity:Mars:545GHz}{1.25}
\setsymbol{HFI:beam:ellipticity:Mars:857GHz}{1.03}

\setsymbol{HFI:CMB:relative:calibration:100GHz}{$\lsim 1\%$}
\setsymbol{HFI:CMB:relative:calibration:143GHz}{$\lsim 1\%$}
\setsymbol{HFI:CMB:relative:calibration:217GHz}{$\lsim 1\%$}
\setsymbol{HFI:CMB:relative:calibration:353GHz}{$\lsim 1\%$}
\setsymbol{HFI:CMB:relative:calibration:545GHz}{}
\setsymbol{HFI:CMB:relative:calibration:857GHz}{}

\setsymbol{HFI:CMB:absolute:calibration:100GHz}{$\lsim 2\%$}
\setsymbol{HFI:CMB:absolute:calibration:143GHz}{$\lsim 2\%$}
\setsymbol{HFI:CMB:absolute:calibration:217GHz}{$\lsim 2\%$}
\setsymbol{HFI:CMB:absolute:calibration:353GHz}{$\lsim 2\%$}
\setsymbol{HFI:CMB:absolute:calibration:545GHz}{}
\setsymbol{HFI:CMB:absolute:calibration:857GHz}{}

\setsymbol{HFI:FIRAS:gain:calibration:accuracy:statistical:100GHz}{}
\setsymbol{HFI:FIRAS:gain:calibration:accuracy:statistical:143GHz}{}
\setsymbol{HFI:FIRAS:gain:calibration:accuracy:statistical:217GHz}{}
\setsymbol{HFI:FIRAS:gain:calibration:accuracy:statistical:353GHz}{2.5\%}
\setsymbol{HFI:FIRAS:gain:calibration:accuracy:statistical:545GHz}{1\%}
\setsymbol{HFI:FIRAS:gain:calibration:accuracy:statistical:857GHz}{0.5\%}

\setsymbol{HFI:FIRAS:gain:calibration:accuracy:systematic:100GHz}{}
\setsymbol{HFI:FIRAS:gain:calibration:accuracy:systematic:143GHz}{}
\setsymbol{HFI:FIRAS:gain:calibration:accuracy:systematic:217GHz}{}
\setsymbol{HFI:FIRAS:gain:calibration:accuracy:systematic:353GHz}{}
\setsymbol{HFI:FIRAS:gain:calibration:accuracy:systematic:545GHz}{7\%}
\setsymbol{HFI:FIRAS:gain:calibration:accuracy:systematic:857GHz}{7\%}

\setsymbol{HFI:FIRAS:zero:point:accuracy:100GHz:units}{0.8\MJysr}
\setsymbol{HFI:FIRAS:zero:point:accuracy:143GHz:units}{}
\setsymbol{HFI:FIRAS:zero:point:accuracy:217GHz:units}{}
\setsymbol{HFI:FIRAS:zero:point:accuracy:353GHz:units}{1.4\MJysr}
\setsymbol{HFI:FIRAS:zero:point:accuracy:545GHz:units}{2.2\MJysr}
\setsymbol{HFI:FIRAS:zero:point:accuracy:857GHz:units}{1.7\MJysr}

\setsymbol{HFI:FIRAS:zero:point:accuracy:100GHz}{0.8}
\setsymbol{HFI:FIRAS:zero:point:accuracy:143GHz}{}
\setsymbol{HFI:FIRAS:zero:point:accuracy:217GHz}{}
\setsymbol{HFI:FIRAS:zero:point:accuracy:353GHz}{1.4}
\setsymbol{HFI:FIRAS:zero:point:accuracy:545GHz}{2.2}
\setsymbol{HFI:FIRAS:zero:point:accuracy:857GHz}{1.7}

\setsymbol{HFI:unit:conversion:100GHz:units}{0.2415\MJysrmK}
\setsymbol{HFI:unit:conversion:143GHz:units}{0.3694\MJysrmK}
\setsymbol{HFI:unit:conversion:217GHz:units}{0.4811\MJysrmK}
\setsymbol{HFI:unit:conversion:353GHz:units}{0.2883\MJysrmK}
\setsymbol{HFI:unit:conversion:545GHz:units}{0.05826\MJysrmK}
\setsymbol{HFI:unit:conversion:857GHz:units}{0.002238\MJysrmK}

\setsymbol{HFI:unit:conversion:100GHz}{0.2415}
\setsymbol{HFI:unit:conversion:143GHz}{0.3694}
\setsymbol{HFI:unit:conversion:217GHz}{0.4811}
\setsymbol{HFI:unit:conversion:353GHz}{0.2883}
\setsymbol{HFI:unit:conversion:545GHz}{0.05826}
\setsymbol{HFI:unit:conversion:857GHz}{0.002238}

\setsymbol{HFI:colour:correction:alpha=-2:V1.01:100GHz}{0.9893}
\setsymbol{HFI:colour:correction:alpha=-2:V1.01:143GHz}{0.9759}
\setsymbol{HFI:colour:correction:alpha=-2:V1.01:217GHz}{1.0007}
\setsymbol{HFI:colour:correction:alpha=-2:V1.01:353GHz}{1.0028}
\setsymbol{HFI:colour:correction:alpha=-2:V1.01:545GHz}{1.0019}
\setsymbol{HFI:colour:correction:alpha=-2:V1.01:857GHz}{0.9889}

\setsymbol{HFI:colour:correction:alpha=0:V1.01:100GHz}{1.0008}
\setsymbol{HFI:colour:correction:alpha=0:V1.01:143GHz}{1.0148}
\setsymbol{HFI:colour:correction:alpha=0:V1.01:217GHz}{0.9909}
\setsymbol{HFI:colour:correction:alpha=0:V1.01:353GHz}{0.9888}
\setsymbol{HFI:colour:correction:alpha=0:V1.01:545GHz}{0.9878}
\setsymbol{HFI:colour:correction:alpha=0:V1.01:857GHz}{1.0014}

%% file: Proj_Ref_6_4a_authors_and_institutes.tex
\author{\small
Planck Collaboration:
P.~A.~R.~Ade\inst{67}
\and
N.~Aghanim\inst{44}
\and
M.~Arnaud\inst{54}
\and
M.~Ashdown\inst{52, 72}
\and
J.~Aumont\inst{44}
\and
C.~Baccigalupi\inst{65}
\and
A.~Balbi\inst{25}
\and
A.~J.~Banday\inst{70, 6, 59}
\and
R.~B.~Barreiro\inst{49}
\and
J.~G.~Bartlett\inst{3, 50}
% ...
}
\institute{\small
Aalto University Mets\"{a}hovi Radio Observatory, Mets\"{a}hovintie 114, FIN-02540 Kylm\"{a}l\"{a}, Finland\\
\and
Agenzia Spaziale Italiana Science Data Center, c/o ESRIN, via Galileo Galilei, Frascati, Italy\\
\and
Astroparticule et Cosmologie, CNRS (UMR7164), Universit\'{e} Denis Diderot Paris 7, B\^{a}timent Condorcet, 10 rue A. Domon et L\'{e}onie Duquet, Paris, France\\
\and
Atacama Large Millimeter/submillimeter Array, ALMA Santiago Central Offices Alonso de Cordova 3107, Vitacura, Casilla 763 0355, Santiago, Chile\\
\and
CITA, University of Toronto, 60 St. George St., Toronto, ON M5S 3H8, Canada\\
\and
CNRS, IRAP, 9 Av. colonel Roche, BP 44346, F-31028 Toulouse cedex 4, France\\
\and
California Institute of Technology, Pasadena, California, U.S.A.\\
\and
DAMTP, Centre for Mathematical Sciences, Wilberforce Road, Cambridge CB3 0WA, U.K.\\
\and
DSM/Irfu/SPP, CEA-Saclay, F-91191 Gif-sur-Yvette Cedex, France\\
\and
DTU Space, National Space Institute, Juliane Mariesvej 30, Copenhagen, Denmark\\
\and
Departamento de F\'{\i}sica, Universidad de Oviedo, Avda. Calvo Sotelo s/n, Oviedo, Spain\\
\and
Department of Astronomy and Astrophysics, University of Toronto, 50 Saint George Street, Toronto, Ontario, Canada\\
\and
Department of Physics \& Astronomy, University of British Columbia, 6224 Agricultural Road, Vancouver, British Columbia, Canada\\
\and
Department of Physics, Gustaf H\"{a}llstr\"{o}min katu 2a, University of Helsinki, Helsinki, Finland\\
\and
Department of Physics, Princeton University, Princeton, New Jersey, U.S.A.\\
\and
Department of Physics, Purdue University, 525 Northwestern Avenue, West Lafayette, Indiana, U.S.A.\\
\and
Department of Physics, University of California, Berkeley, California, U.S.A.\\
\and
Department of Physics, University of California, One Shields Avenue, Davis, California, U.S.A.\\
\and
Department of Physics, University of California, Santa Barbara, California, U.S.A.\\
\and
Department of Physics, University of Illinois at Urbana-Champaign, 1110 West Green Street, Urbana, Illinois, U.S.A.\\
\and
Dipartimento di Fisica G. Galilei, Universit\`{a} degli Studi di Padova, via Marzolo 8, 35131 Padova, Italy\\
\and
Dipartimento di Fisica, Universit\`{a} La Sapienza, P. le A. Moro 2, Roma, Italy\\
\and
Dipartimento di Fisica, Universit\`{a} degli Studi di Milano, Via Celoria, 16, Milano, Italy\\
\and
Dipartimento di Fisica, Universit\`{a} degli Studi di Trieste, via A. Valerio 2, Trieste, Italy\\
\and
Dipartimento di Fisica, Universit\`{a} di Roma Tor Vergata, Via della Ricerca Scientifica, 1, Roma, Italy\\
\and
Discovery Center, Niels Bohr Institute, Blegdamsvej 17, Copenhagen, Denmark\\
\and
Dpto. Astrof\'{i}sica, Universidad de La Laguna (ULL), E-38206 La Laguna, Tenerife, Spain\\
\and
European Southern Observatory, ESO Vitacura, Alonso de Cordova 3107, Vitacura, Casilla 19001, Santiago, Chile\\
\and
European Space Agency, ESAC, Planck Science Office, Camino bajo del Castillo, s/n, Urbanizaci\'{o}n Villafranca del Castillo, Villanueva de la Ca\~{n}ada, Madrid, Spain\\
\and
European Space Agency, ESTEC, Keplerlaan 1, 2201 AZ Noordwijk, The Netherlands\\
\and
Haverford College Astronomy Department, 370 Lancaster Avenue, Haverford, Pennsylvania, U.S.A.\\
\and
Helsinki Institute of Physics, Gustaf H\"{a}llstr\"{o}min katu 2, University of Helsinki, Helsinki, Finland\\
\and
INAF - Osservatorio Astrofisico di Catania, Via S. Sofia 78, Catania, Italy\\
\and
INAF - Osservatorio Astronomico di Padova, Vicolo dell'Osservatorio 5, Padova, Italy\\
\and
INAF - Osservatorio Astronomico di Roma, via di Frascati 33, Monte Porzio Catone, Italy\\
\and
INAF - Osservatorio Astronomico di Trieste, Via G.B. Tiepolo 11, Trieste, Italy\\
\and
INAF/IASF Bologna, Via Gobetti 101, Bologna, Italy\\
\and
INAF/IASF Milano, Via E. Bassini 15, Milano, Italy\\
\and
INRIA, Laboratoire de Recherche en Informatique, Universit\'{e} Paris-Sud 11, B\^{a}timent 490, 91405 Orsay Cedex, France\\
\and
IPAG: Institut de Plan\'{e}tologie et d'Astrophysique de Grenoble, Universit\'{e} Joseph Fourier, Grenoble 1 / CNRS-INSU, UMR 5274, Grenoble, F-38041, France\\
\and
ISDC Data Centre for Astrophysics, University of Geneva, ch. d'Ecogia 16, Versoix, Switzerland\\
\and
Imperial College London, Astrophysics group, Blackett Laboratory, Prince Consort Road, London, SW7 2AZ, U.K.\\
\and
Infrared Processing and Analysis Center, California Institute of Technology, Pasadena, CA 91125, U.S.A.\\
\and
Institut d'Astrophysique Spatiale, CNRS (UMR8617) Universit\'{e} Paris-Sud 11, B\^{a}timent 121, Orsay, France\\
\and
Institut d'Astrophysique de Paris, CNRS UMR7095, Universit\'{e} Pierre \& Marie Curie, 98 bis boulevard Arago, Paris, France\\
\and
Institute of Astronomy and Astrophysics, Academia Sinica, Taipei, Taiwan\\
\and
Institute of Theoretical Astrophysics, University of Oslo, Blindern, Oslo, Norway\\
\and
Instituto de Astrof\'{\i}sica de Canarias, C/V\'{\i}a L\'{a}ctea s/n, La Laguna, Tenerife, Spain\\
\and
Instituto de F\'{\i}sica de Cantabria (CSIC-Universidad de Cantabria), Avda. de los Castros s/n, Santander, Spain\\
\and
Jet Propulsion Laboratory, California Institute of Technology, 4800 Oak Grove Drive, Pasadena, California, U.S.A.\\
\and
Jodrell Bank Centre for Astrophysics, Alan Turing Building, School of Physics and Astronomy, The University of Manchester, Oxford Road, Manchester, M13 9PL, U.K.\\
\and
Kavli Institute for Cosmology Cambridge, Madingley Road, Cambridge, CB3 0HA, U.K.\\
\and
LERMA, CNRS, Observatoire de Paris, 61 Avenue de l'Observatoire, Paris, France\\
\and
Laboratoire AIM, IRFU/Service d'Astrophysique - CEA/DSM - CNRS - Universit\'{e} Paris Diderot, B\^{a}t. 709, CEA-Saclay, F-91191 Gif-sur-Yvette Cedex, France\\
\and
Laboratoire Traitement et Communication de l'Information, CNRS (UMR 5141) and T\'{e}l\'{e}com ParisTech, 46 rue Barrault F-75634 Paris Cedex 13, France\\
\and
Laboratoire de Physique Subatomique et de Cosmologie, CNRS, Universit\'{e} Joseph Fourier Grenoble I, 53 rue des Martyrs, Grenoble, France\\
\and
Laboratoire de l'Acc\'{e}l\'{e}rateur Lin\'{e}aire, Universit\'{e} Paris-Sud 11, CNRS/IN2P3, Orsay, France\\
\and
Lawrence Berkeley National Laboratory, Berkeley, California, U.S.A.\\
\and
Max-Planck-Institut f\"{u}r Astrophysik, Karl-Schwarzschild-Str. 1, 85741 Garching, Germany\\
\and
MilliLab, VTT Technical Research Centre of Finland, Tietotie 3, Espoo, Finland\\
\and
National University of Ireland, Department of Experimental Physics, Maynooth, Co. Kildare, Ireland\\
\and
Niels Bohr Institute, Blegdamsvej 17, Copenhagen, Denmark\\
\and
Observational Cosmology, Mail Stop 367-17, California Institute of Technology, Pasadena, CA, 91125, U.S.A.\\
\and
Optical Science Laboratory, University College London, Gower Street, London, U.K.\\
\and
SISSA, Astrophysics Sector, via Bonomea 265, 34136, Trieste, Italy\\
\and
SUPA, Institute for Astronomy, University of Edinburgh, Royal Observatory, Blackford Hill, Edinburgh EH9 3HJ, U.K.\\
\and
School of Physics and Astronomy, Cardiff University, Queens Buildings, The Parade, Cardiff, CF24 3AA, U.K.\\
\and
Space Sciences Laboratory, University of California, Berkeley, California, U.S.A.\\
\and
Stanford University, Dept of Physics, Varian Physics Bldg, 382 Via Pueblo Mall, Stanford, California, U.S.A.\\
\and
Universit\'{e} de Toulouse, UPS-OMP, IRAP, F-31028 Toulouse cedex 4, France\\
\and
Universities Space Research Association, Stratospheric Observatory for Infrared Astronomy, MS 211-3, Moffett Field, CA 94035, U.S.A.\\
\and
University of Cambridge, Cavendish Laboratory, Astrophysics group, J J Thomson Avenue, Cambridge, U.K.\\
% ...
}